\documentclass[
reprint,
superscriptaddress,
 amsmath,
 amssymb,
 amsfonts,
 aps,
prb,
]{revtex4-2}

\usepackage{graphicx}
\usepackage{dcolumn}
\usepackage{bm}
\usepackage{hyperref}
\usepackage[english]{babel}
\usepackage[subpreambles=false]{standalone}
\usepackage{import}

\begin{document}

\title{Local imaging of diamagnetism in proximity coupled niobium nano-island arrays on gold thin films}

\author{Logan Bishop-Van Horn}
\altaffiliation{Contributed equally to this work}
\affiliation{Department of Physics, Stanford University, Stanford, California 94305, USA}
\affiliation{Stanford Institute for Materials and Energy Sciences, SLAC National Accelerator Laboratory, Menlo Park, California 94025, USA}

\author{Irene P. Zhang}
\altaffiliation{Contributed equally to this work}
\affiliation{Department of Applied Physics, Stanford University, Stanford, California 94305, USA}
\affiliation{Stanford Institute for Materials and Energy Sciences, SLAC National Accelerator Laboratory, Menlo Park, California 94025, USA}

\author{Emily N. Waite}
\affiliation{Department of Physics and Materials Research Laboratory, University of Illinois, Urbana, Illinois 61801, USA}

\author{Ian Mondragon-Shem}
\affiliation{Department of Physics and Astronomy, Northwestern University, Evanston, Illinois 60208, USA}
\affiliation{Materials Science Division, Argonne National Laboratory, Argonne, Illinois 60439, USA}

\author{Scott Jensen}
\affiliation{Department of Physics and Materials Research Laboratory, University of Illinois, Urbana, Illinois 61801, USA}

\author{Junseok Oh}
\affiliation{Department of Physics and Materials Research Laboratory, University of Illinois, Urbana, Illinois 61801, USA}

\author{Tom Lippman}
\affiliation{Department of Physics, Stanford University, Stanford, California 94305, USA}

\author{Malcolm Durkin}
\affiliation{Department of Physics, University of Colorado Boulder, Boulder, Colorado 80305, USA}

\author{Taylor L. Hughes}
\affiliation{Department of Physics and Materials Research Laboratory, University of Illinois, Urbana, Illinois 61801, USA}

\author{Nadya Mason}
\affiliation{Department of Physics and Materials Research Laboratory, University of Illinois, Urbana, Illinois 61801, USA}

\author{Kathryn A. Moler}
\affiliation{Department of Applied Physics, Stanford University, Stanford, California 94305, USA}
\affiliation{Stanford Institute for Materials and Energy Sciences, SLAC National Accelerator Laboratory, Menlo Park, California 94025, USA}

\author{Ilya Sochnikov}
\email{ilya.sochnikov@uconn.edu}
\affiliation{Department of Physics, University of Connecticut, Storrs, Connecticut 06269, USA}
\affiliation{Institute of Material Science, University of Connecticut, Storrs, Connecticut 06269, USA}

\date{\today}

\begin{abstract}

In this work we study the effect of engineered disorder on the local magnetic response of proximity coupled superconducting island arrays by comparing scanning Superconducting Quantum Interference Device (SQUID) susceptibility measurements to a model in which we treat the system as a network of one-dimensional (1D) superconductor-normal metal-superconductor (SNS) Josephson junctions, each with a Josephson coupling energy $E_J$ determined by the junction length or distance between islands. We find that the disordered arrays exhibit a spatially inhomogeneous diamagnetic response which, for low local applied magnetic fields, is well described by this junction network model, and we discuss these results as they relate to inhomogeneous two-dimensional (2D) superconductors. Our model of the static magnetic response of the arrays does not fully capture the onset of nonlinearity and dissipation with increasing applied field, as these effects are associated with vortex motion due to the dynamic nature of the scanning SQUID susceptometry measurement. This work demonstrates a model 2D superconducting system with engineered disorder and highlights the impact of dissipation on the local magnetic properties of 2D superconductors and Josephson junction arrays.

\end{abstract}

\maketitle

\section{Introduction}

The effects of disorder and inhomogeneity on the superconducting properties of thin films have attracted both practical and theoretical interest, particularly in the presence of spatial correlations~\cite{alexander_1983,gastiasoro_2016,lippman_agreement_2012}. Disorder has been shown to weaken superconductivity in  single crystal $\text{Sr}_2\text{RuO}_4$ and thin film $\text{YBa}_2\text{Cu}_2\text{O}_{7-\delta}$~\cite{mackenzie_1998,ye_1994}. In contrast, disorder in monolayers of $\text{TaS}_2$ was recently reported to enhance the critical transition temperature~\cite{peng_2018}. In single atomic layers of lead on silicon, the presence of disorder can lead to atomically short Josephson weak links~\cite{brun_2014}. The addition of correlated  disorder in superconductors can be beneficial for some aspects of superconductivity compared to systems without correlation~\cite{alloul_2009}. In $\text{YBa}_2\text{Cu}_2\text{O}_{7-\delta},$ microstructures of columnar defects introduced by irradiation pin flux lines more strongly than random point defects, shifting the irreversibility line significantly upward~\cite{civale_1991,nelson_1992,tesanovic_1994}. In two-dimensional (2D) superconductor-to-insulator systems, the existence of disorder is thought to create weakly coupled islands of superconductivity; with increasing disorder, such islands can remain even beyond the superconductor-insulator transition~\cite{dubi_2007,crane_2007}. 
In numerical studies of the quantum XY model, correlated disorder was shown to cause a broadening of the Berezinskii-Kosterlitz-Thouless (BKT) transition with respect to temperature, whereas uncorrelated disorder had no effect on the sharpness of the transition~\cite{maccari_2017}. Inhomogeneity at any length scale can also complicate interpretation of bulk or sample-averaged measurements.

Given the effect of both correlated and uncorrelated disorder on superconducting systems, one open question is: how do spatial correlations in disorder affect the magnetic response of a 2D superconducting system? To explore this question, we used scanning Superconducting Quantum Interference Device microscopy (scanning SQUID microscopy or SSM) to measure the local diamagnetic response of arrays of niobium islands with proximity coupling via a thin layer of gold. Superconducting island arrays on normal metal can be a useful model for studying disorder in 2D superconductors, as both the disorder and critical current can be engineered by changing the spacing between superconducting islands~\cite{spivak_1982, eley_approaching_2012,Naibert2021}.

Previously, Eley \textit{et al.} showed that transport in ordered arrays of niobium islands on gold can be modeled by treating the entire array as a single diffusive superconductor-normal metal-superconductor (SNS) junction with a junction  length equal to the island-to-island spacing, $d$~\cite{eley_dependence_2013}. Neighboring islands couple to each other by proximitizing the underlying gold, and the system undergoes a BKT transition with decreasing temperature~\cite{berezinskit_1972,kosterlitz_1973, Han2014}. In this work we study the effect of engineered disorder on the local magnetic response of proximity coupled superconducting island arrays by comparing SSM susceptibility measurements to a model in which we treat the system as a network of one-dimensional (1D) SNS Josephson junctions, each with a Josephson coupling energy $E_J$ determined by the junction length or distance between islands.
We find that the disordered arrays exhibit a spatially inhomogeneous diamagnetic response which, for low applied magnetic fields, is well described by this junction network model. Upon increasing the applied field, the response becomes nonlinear and dissipative, with both the degree of nonlinearity and the spatial structure of the dissipative effects depending strongly on the details of the engineered disorder in a way that is not fully captured by the model.

\section{Methods}
\begin{figure}
    \centering
    \includegraphics[width=\linewidth]{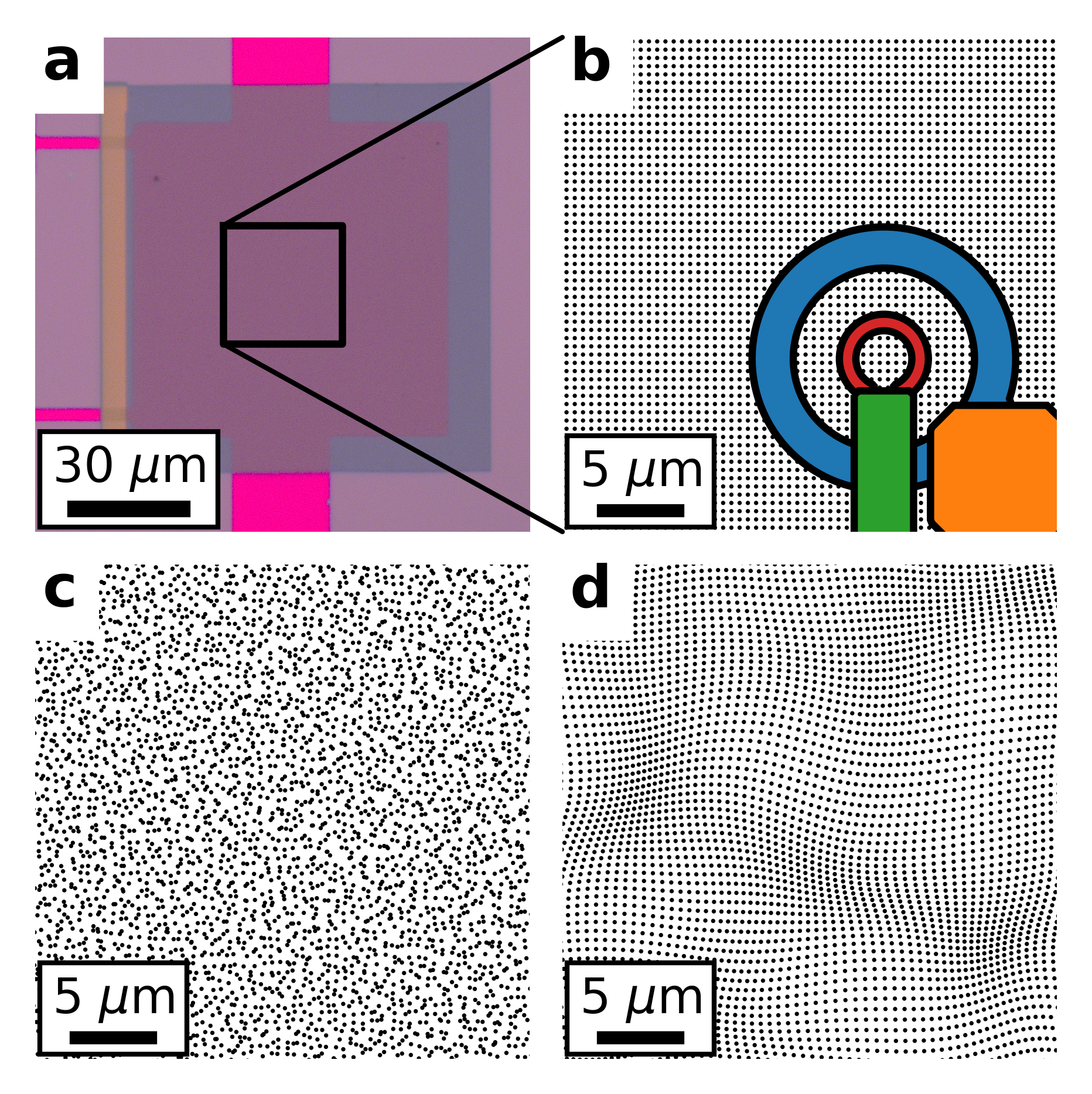}
    \caption{Device and sensor geometry. (a) Optical image of the ordered array. The background is the SiO${}_{2}$ substrate, the bright pink regions are bare gold film, the gray region is niobium islands directly on SiO${}_{2}$, and the dark pink central $80\,\mu\mathrm{m}\times80\,\mu\mathrm{m}$ region is niobium islands on top of the gold film. (b-d) Schematic of the designed island positions for a $30\,\mu\mathrm{m}\times 30\,\mu\mathrm{m}$ region of (b) the ordered array, (c) an uncorrelated disordered array ($\sigma=40\%$), and (d) a correlated disordered array ($\sigma=15\%$). The islands are drawn to scale, i.e., each island has a diameter of 260 nm. The SQUID susceptometer geometry is shown in (b), with the field coil in blue (large radius) and the pickup loop in red (small radius). There are also superconducting shields covering the field coil and pickup loop leads, which are shown as orange and green polygons, respectively.}
    \label{fig:islands-squid}
\end{figure}

We measured arrays with three types of island configurations: 1) ordered, 2) uncorrelated disorder, and 3) correlated disorder (Figure~\ref{fig:islands-squid}). The ordered array consists of a $100\times 100\,\mu\mathrm{m}^2$ square lattice of nominally circular niobium islands, with a lattice constant $a = 500$ nm; each island in the array has a diameter $D_\mathrm{island}=260\,\mathrm{nm}$ and a height of 100 nm. The minimum edge-to-edge island spacing in the ordered array is therefore $d_0=a-D_\mathrm{island}=240\,\mathrm{nm}$. The disordered arrays were generated by displacing the island positions in the ordered array by a distance $\vert\Delta \textbf{R} \vert$ drawn from a normal distribution $f(\vert\Delta \textbf{R} \vert) = \frac{1}{\sigma\sqrt{2\pi}}\exp{\left(-\frac{\vert\Delta \textbf{R} \vert^2}{2\sigma^2}\right)}$  with standard deviation $\sigma$~\cite{lippman_2013}.
For arrays with uncorrelated disorder, this process defined the final island locations. To generate correlated disorder, an additional filter kernel was applied to normally distributed island displacements so that the covariance of the displacements of any two islands is described by the correlation function
\begin{equation}
    R(z) = \sigma^2\exp{\left(-\frac{z^2}{\ell^2}\right)},
\end{equation}
where $z$ is the center-to-center distance between pairs of islands in the square lattice and $\ell= 5\,\mu\mathrm{m}$ is the correlation length~\cite{lippman_2013}.

For each array, the locations of the islands were patterned onto a 10 nm thick gold film with lateral dimensions $80\times80$ $\mu\mathrm{m}^2$ using electron beam lithography, after which the niobium islands were deposited using electron beam evaporation. Transport measurements performed in a dilution refrigerator with a base temperature of 10 mK show two distinct transition temperatures: one from the niobium islands themselves entering the superconducting regime ($T_1$) and a second, lower transition temperature ($T_2$) marking the onset of phase coherence in the proximitized gold film~\cite{eley_approaching_2012} (see Supplemental Material~\cite{supp}, which includes Refs.~\cite{Gardner2001-gr,kirtley_scanning_2016,kirtley_response_2016,Kogan2011-zn,Cave1986-js,brandt_thin_2005,Jackson1999-zb,Paton1969-kj,Shewchuk1996-va,Bishop-Van_Horn2022-sy}). The superconducting transition temperature of the niobium islands themselves is lower than that of bulk niobium, consistent with previous work~\cite{durkin_2020}.

The arrays were studied using a scanning SQUID microscope mounted in a helium-3 refrigerator at its base temperature, $T=400\,\mathrm{mK}$. The scanning SQUID susceptometer used in this work consists of gradiometric concentric pickup loop and field coil pairs, with pickup loop inner radius $r^\mathrm{inner}_\mathrm{PL}=1.7\,\mu\mathrm{m}$ (outer radius $r^\mathrm{outer}_\mathrm{PL}=2.7\,\mu\mathrm{m}$) and field coil inner radius $r^\mathrm{inner}_\mathrm{FC}=5.5\,\mu\mathrm{m}$ (outer radius $r^\mathrm{outer}_\mathrm{FC}=8.0\,\mu\mathrm{m}$) (Fig.~\ref{fig:islands-squid} (b))~\cite{huber2008gradiometric}. Using an SR830 lock-in amplifier, we apply a local low-frequency AC magnetic field to the array using the field coil carrying current $I_\mathrm{FC}$ and record the flux $\Phi_\mathrm{PL}$ through the pickup loop generated by the array's response as a function of the field coil position. We define this response, normalized by the current through the field coil, as the local susceptibility, $\phi=\Phi_\mathrm{PL}/I_\mathrm{FC}$, which we report in units of $\Phi_0/\mathrm{A}$ where $\Phi_0$ is the superconducting flux quantum. The gradiometric design of the SQUID susceptometer allows us to detect a local magnetic response due to current flowing in the arrays, $\Phi_\mathrm{PL}=\phi I_\mathrm{FC}$, which is much smaller than the flux through the pickup loop due to the field coil. For example, the measurements presented below have a typical signal magnitude of $\sim1\,\Phi_0/\mathrm{A}$ (with signal-to-noise ratio $\gg 1$ except at very small $I_\mathrm{FC}$), which is approximately 1/1000 of the intrinsic mutual inductance between the field coil and pickup loop. (See Supplemental Material~\cite{supp} for a more detailed description of the sensor design and susceptometry measurement technique.) The susceptibility $\phi$ has a component $\phi'$ that is in-phase with the applied field and a component $\phi''$ that is out-of-phase with the applied field. The in-phase component is a measure of the local diamagnetic screening in a superconducting sample and hence the London penetration depth or superfluid density, while the out-of-phase component is a measure of dissipative currents. The measurements were performed at nominally zero applied global field, so that the only applied field was the local AC field from the susceptometer field coil.

\section{Results and Discussion}

\begin{table}
    \centering
    \begin{tabular}{c|c|c}
    \hline
        Sample & Type of Disorder & $\sigma$\\
    \hline
        Ordered & Ordered & 0\% \\
        UC3 & Uncorrelated & 40\% \\
        UC4 & Uncorrelated & Uniformly distributed \\
        C2 & Correlated & 10\% \\
        C3 & Correlated & 15\% \\
    \hline
    \end{tabular}
    \caption{Designed disorder in measured arrays. The standard deviation $\sigma$ of the island displacement distribution is expressed as a percentage of the island spacing in the ordered array. The island positions in UC4 are generated without reference to the ordered array. Instead, we randomly select points in a $100\times100\,\mu\mathrm{m}^2$ area according to a uniform distribution.}
    \label{tbl:samples}
\end{table}

\begin{figure*}
  \includegraphics[width=\textwidth]{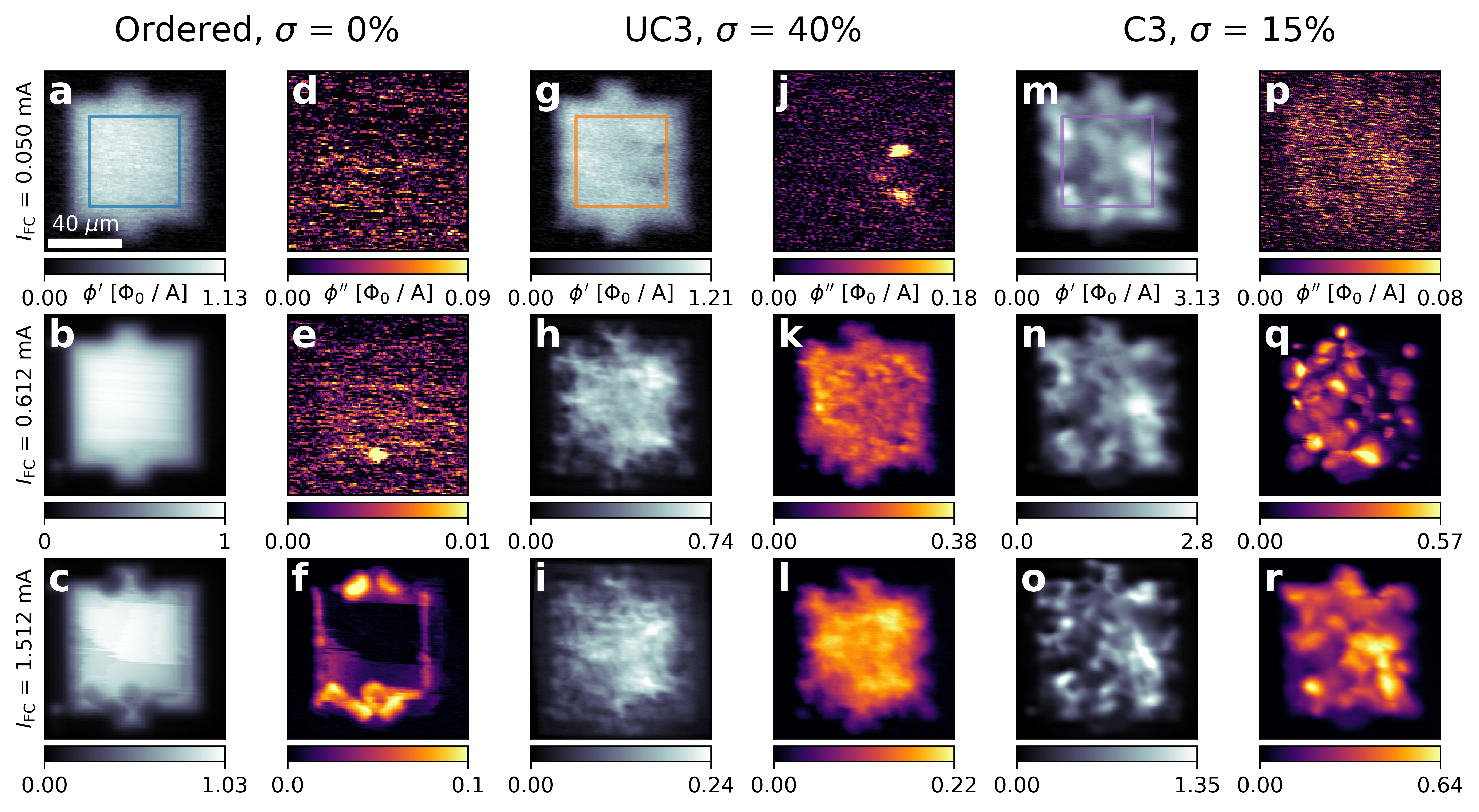}
  \caption{Inhomogeneous, nonlinear diamagnetic response in disordered arrays under increasing local applied field. In-phase susceptibility $\phi'$ (black-to-white colormap) and out-of-phase susceptibility $\phi''$ (black-to-yellow colormap) for the three arrays whose island positions are shown in Figure~\ref{fig:islands-squid} (b-d), measured at three different RMS field coil currents $I_\mathrm{FC}$. All images share the scale bar given in panel (a) and were taken at the same nominal sensor height and temperature. Note that panels (d, e, j, p) only show noise, as $\phi''$ was below the measurement noise floor for these measurements with small $I_\mathrm{FC}$. The colored boxes in panels (a, g, m) indicate the central $50\times50\,\mu\mathrm{m}^2$ regions for which the distributions of $\phi'$ are shown as a function of $I_\mathrm{FC}$ in Figure~\ref{fgr:SSMdists} (a, b, e), respectively. Measurements of all five samples listed in Table~\ref{tbl:samples} can be found in the Supplemental Material~\cite{supp}.
}
\label{fgr:SSMdata}
\end{figure*}

We imaged five arrays: one completely ordered, two with uncorrelated disorder, and two with correlated disorder, as summarized in Table \ref{tbl:samples}. We applied root-mean-square (RMS) field coil currents $I_\mathrm{FC}$ from 0.012 mA up to 3.024 mA, corresponding to about 1 $\mu$T to 300 $\mu$T at the center of the field coil, at a frequency of $f_\mathrm{FC}=800\,\mathrm{Hz}$ to probe the linear and nonlinear regimes of the diamagnetic response. The in-phase susceptibility images reveal striking differences in the spatial structure of the diamagnetic susceptibility between the completely ordered array and arrays with disorder (Figure~\ref{fgr:SSMdata}). At the lowest field coil current, the ordered array shows homogeneous diamagnetic screening (Figure~\ref{fgr:SSMdata} (a)). From the magnitude of the in-phase susceptibility signal, we estimate an effective 2D penetration depth (equal to half the Pearl length~\cite{Pearl1964-cl}) of $\Lambda> 500\,\mu\mathrm{m}$, indicating weak Meissner screening and a small superfluid density~\cite{Kirtley2012-od}. Only at higher field coil currents does spatial structure appear in the form of reduced diamagnetism at the edge of the array (Figure \ref{fgr:SSMdata} (b-c)). In contrast, in all the arrays with engineered disorder, our measurements reveal significant spatial inhomogeneity in the local diamagnetic response. The diamagnetism in these arrays varies on a length scale of a few microns over the entirety of the array (Figure \ref{fgr:SSMdata} (g-i) and (m-o)). In arrays with correlated disorder, this inhomogeneity can be seen even at the smallest applied field (Figure \ref{fgr:SSMdata} (m)).

\begin{figure*}
  \includegraphics[width=\textwidth]{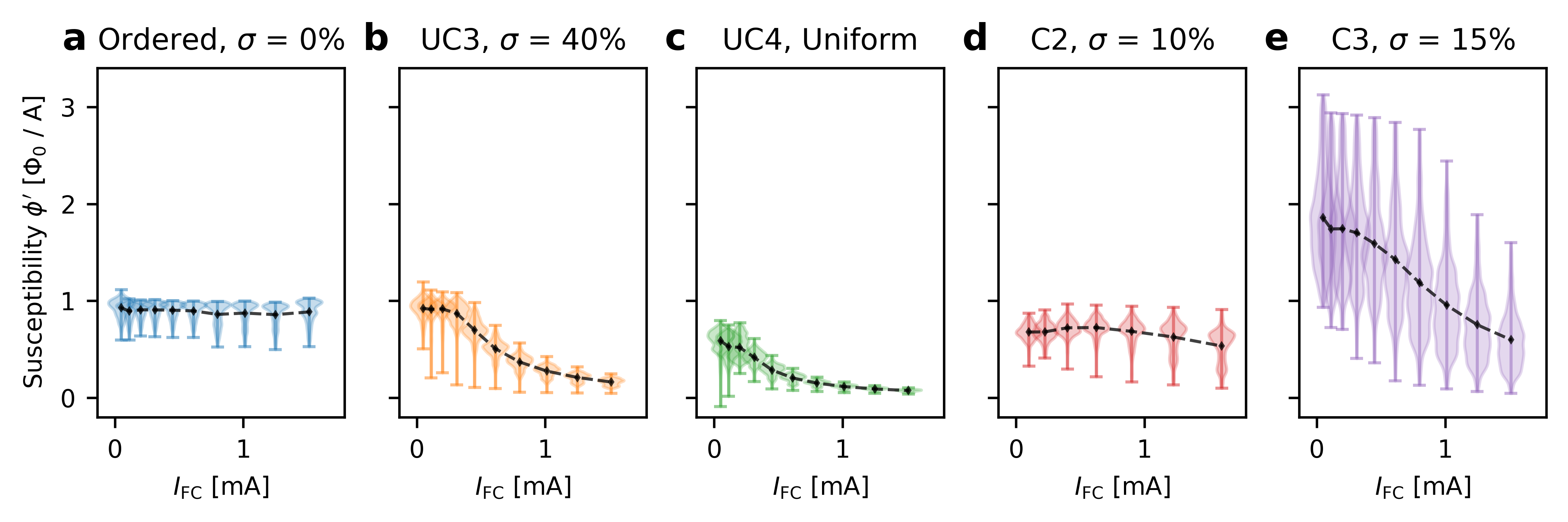}
  \caption{Distribution of in-phase susceptibility $\phi'$ in SSM images for all measured samples as a function of field coil current. The shaded ``violins'' show the empirical probability distribution of the susceptibility signal $\phi'$ in the central $50\times50\,\mu\mathrm{m}^2$ region of each array (see colored boxes in Figure~\ref{fgr:SSMdata} (a, g, m)), and the black diamonds and dashed lines show the average over this region. The response of the ordered array (a) is linear in the sense that the magnitude and spatial distribution of the diamagnetic susceptibility are constant as a function of the applied field. In contrast, the disordered arrays (b-e) respond nonlinearly at all but the lowest field coil currents. The magnitude of $\phi'$ depends strongly on the sensor height, or the distance between the SQUID field coil and the sample. The data for C2 were taken at $2-3\,\mu\mathrm{m}$ higher sensor height compared to the ordered and uncorrelated arrays, resulting in a relative overall reduction in signal magnitude. Similarly, the stronger average in-phase susceptibility for C3 most likely indicates that this sample was measured with a smaller sensor height than the ordered and uncorrelated arrays~\cite{supp}.
}
\label{fgr:SSMdists}
\end{figure*}

The magnitude and spatial structure of the diamagnetic susceptibility is not constant as a function of the applied field. In Figure \ref{fgr:SSMdists}, the distribution of in-phase diamagnetic susceptibility for each array is plotted as a function of applied field coil current, revealing differences between the ordered and disordered samples in the linearity of the diamagnetic response with respect to the local applied field. Because $\phi'$ necessarily goes to zero at the edges of the arrays, in Figure~\ref{fgr:SSMdists} we plot the distribution of $\phi'$ only from the central $50\times50\,\mu\mathrm{m}^2$ region of each array (see colored boxes in Figure~\ref{fgr:SSMdata} (a, g, m)). Except near the edges of the gold film, the ordered sample remains in the linear regime (i.e., the susceptibility is constant as a function of applied field and the out-of-phase susceptibility $\phi''$ is small) up to $I_\mathrm{FC}$ = 1.512 mA (Figure~\ref{fgr:SSMdata} (a-f) and Figure~\ref{fgr:SSMdists} (a)), while the disordered samples enter the nonlinear regime, with decreasing in-phase susceptibility $\phi'$ and significant dissipative out-of-phase susceptibility $\phi''$, at applied field coil currents as low as $I_\mathrm{FC}$ = 0.312 mA (Figure~\ref{fgr:SSMdata} (g-l) and (m-r), Figure~\ref{fgr:SSMdists} (b-e)). At the highest $I_\mathrm{FC}$, the ordered array begins to exhibit an inhomogeneous response (Figure~\ref{fgr:SSMdata} (c, f)), which we attribute to heating of the gold film due to vortex motion near the edges of the array~\cite{supp}.

\begin{figure}
  \includegraphics[width=0.98\linewidth]{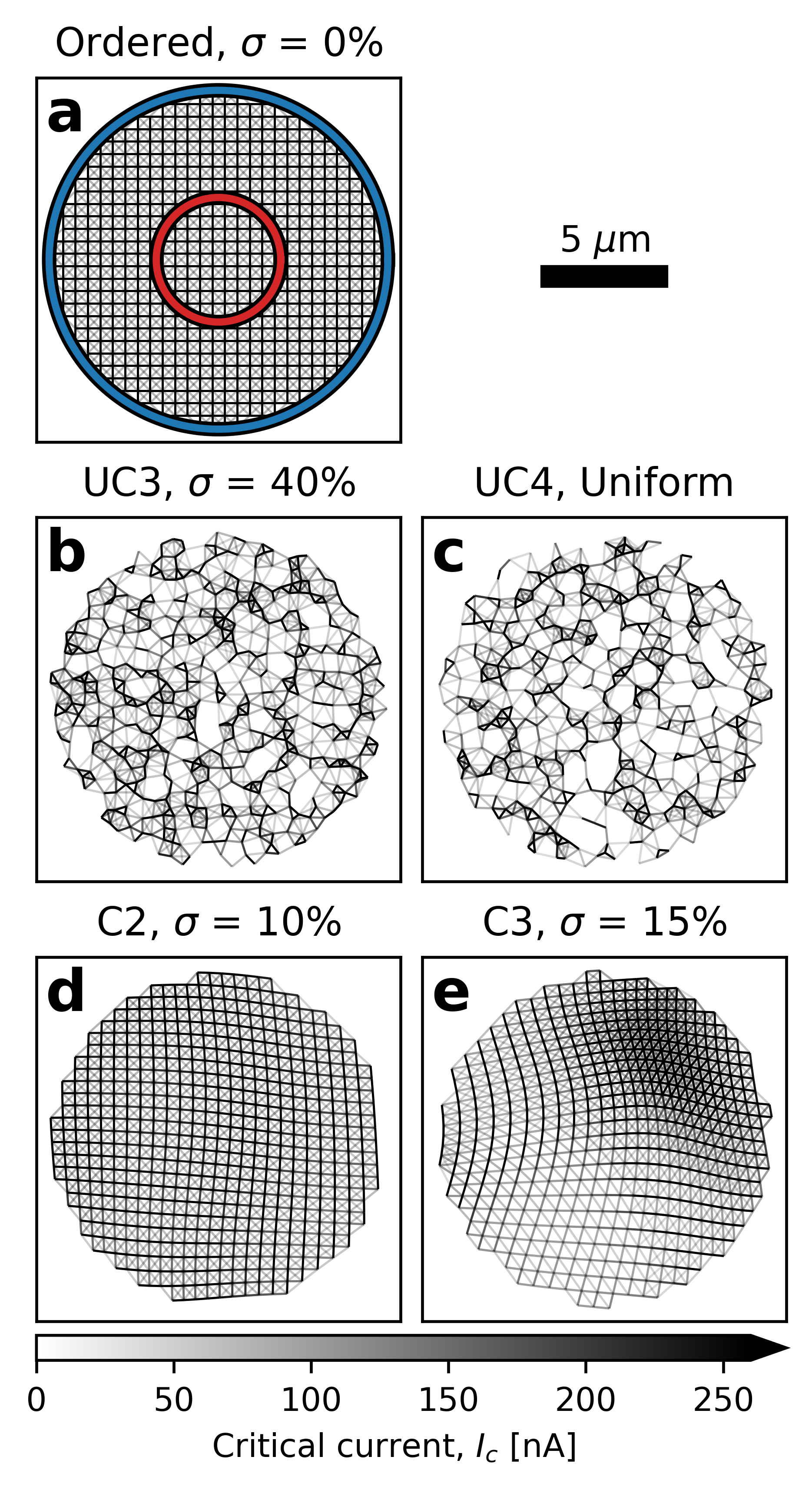}
  \caption{Junction networks for a single field coil position for each of the five arrays studied in this work. Each line represents a single junction with critical current $I_c(d)=I_0(d_0/d)^2$ indicated by the shade of the line as shown by the colorbar. The colorbar is saturated at 260 nA because in the uncorrelated arrays there are a small number of very short (i.e. high critical current) junctions. Locations where lines intersect correspond to the center of a niobium island. For all five arrays, we use a junction cutoff radius $r_\mathrm{cutoff}=0.9\,\mu\mathrm{m}$ and a critical current scale of $I_0=260\,\mathrm{nA}$. For the sake of visibility, the networks shown here use a patch radius $r_\mathrm{patch}=r_\mathrm{FC}=6.8\,\mu\mathrm{m}$, which is a factor of 2.5 smaller than the patch radius used in the simulations. (a) The ordered array: a square lattice of islands with lattice constant $a=500\,\mathrm{nm}$. The blue circle shows the 1D loop used to model the field coil, and the red circle shows the 1D loop with radius $r_\mathrm{PL}=2.5\,\mu\mathrm{m}$ used to model the pickup loop. (b) Array UC3: uncorrelated disorder with $\sigma=40\%$. (c) Array UC1: uniformly distributed island positions. (c) Array C2: correlated disorder with $\sigma=10\%$. (d) Array C3: correlated disorder with $\sigma=15\%$.}
\label{fgr:sim-setup}
\end{figure}

To explore the role of disorder in the inhomogeneous, nonlinear magnetic response of this engineered 2D superconductor, we have modeled the system as a network of 1D SNS Josephson junctions in which pairs of adjacent islands form junctions with critical current $I_c$, or Josephson energy $E_J=I_c\Phi_0/2\pi$, determined by the junction length or edge-to-edge island spacing $d$ (see Figure~\ref{fgr:sim-setup}). Eley \textit{et al.} found that for ordered triangular arrays of niobium nano-islands on gold, the dependence of the critical current of the entire array on edge-to-edge island spacing $d$ and temperature $T$ is well described by the expression for a single diffusive SNS junction:
\begin{equation}
I_c(d, T) = c_0\frac{E_\mathrm{Th} (d)}{eR_N}\left[1-c_1\exp\left(-\frac{c_0E_\mathrm{Th} (d)}{3.2k_\mathrm{B}T}\right)\right],
\label{eq:full_Ic}
\end{equation}
where $R_N$ is the normal state resistance, $k_\mathrm{B}$ is Boltzmann's constant, and $c_0$ and $c_1$ are dimensionless fitting parameters which are of order 1~\cite{eley_dependence_2013,dubos_sns_2001}. The Thouless energy $E_\mathrm{Th}$ is given by $E_\mathrm{Th}(d)=\hbar D/d^2,$ where $D$ is the normal metal diffusion constant, with $D=95\,\mathrm{cm}^2/\mathrm{s}$ for the gold films studied here. At temperatures $T$ that are small compared to the Thouless energy $E_\mathrm{Th}/k_\mathrm{B}$ ($\approx1.26\,\mathrm{K}$ for the ordered array), Equation~\ref{eq:full_Ic} is dominated by the term $c_0E_\mathrm{Th}/eR_N\propto 1/d^2$. We therefore assume that the critical current of each junction is given by $I_c(d)=I_0(d_0/d)^2$, where $d_0=240\,\mathrm{nm}$ is the minimum island spacing for the ordered array (Figure~\ref{fig:islands-squid} (b)) and $I_0$ is a constant that corresponds to the maximum critical current per junction in the ordered array. The value of $I_0$ determines the overall strength of the diamagnetic response and select it to roughly match the magnitude of the measured in-phase susceptibility.

We model the field coil and pickup loop as 1D circular loops with radii $r_\mathrm{FC}=6.8\,\mu\mathrm{m}$ and $r_\mathrm{PL}=2.5\,\mu\mathrm{m}$ respectively (see Figure~\ref{fgr:sim-setup}). Given the applied magnetic vector potential $\mathbf{A}(\mathbf{r})$ due to a current $I_\mathrm{FC}$ in the field coil, we solve for the superconducting phase $\varphi_i$ of each island $i$ centered at position $\mathbf{r}_i$ in the network, subject to the constraints of current conservation and phase single-valuedness, via a large-scale nonlinear programming solver~\cite{supp,Beal2018-dq,Hedengren2014-zu}. We then calculate the  supercurrent flowing between each pair of islands $(i, j)$ assuming a sinusoidal current-phase relation $I_{ij}(\theta_{ij}, d_{ij})=I_c(d_{ij})\sin\theta_{ij}$, where $d_{ij}=|\mathbf{r}_i-\mathbf{r}_j|-D_\mathrm{island}$ is the junction length and $\theta_{ij}=\varphi_j-\varphi_i-2\pi\Phi_0^{-1}\int_{\mathbf{r}_i}^{\mathbf{r}_j}\mathbf{A}(\mathbf{r})\cdot\mathrm{d}\mathbf{r}$ is the gauge-invariant phase across the junction. Finally, we compute the flux $\Phi_\mathrm{PL}$ through the pickup loop due to the supercurrent flowing in the network to obtain a simulated in-phase susceptibility $\phi'=\Phi_\mathrm{PL}/I_\mathrm{FC}$.

It is not computationally practical to model the response of all  $\sim25,000$ islands in an array simultaneously, so we make two simplifying approximations. First, for each field coil position, we construct a network or graph containing only islands inside a ``patch'' of radius $r_\mathrm{patch}=17\,\mu\mathrm{m}=2.5\times r_\mathrm{FC}$ around the center of the field coil. If the field coil is positioned $z_\mathrm{FC}=2\,\mu\mathrm{m}$ above the array, the magnitude of the field from the field coil at the edge of patch is $\approx4\%$ of the field at the center and junctions outside of the patch are at least $r_\mathrm{patch}=6.8\times r_\mathrm{PL}$ away from the center of the pickup loop. Therefore, junctions outside of $r_\mathrm{patch}$ are both weakly influenced by the field from the field coil and inefficient at coupling flux into the pickup loop, such that they don't contribute significantly to the susceptibility signal. In practice, increasing $r_\mathrm{patch}$ by 20\%, from $2.5\times r_\mathrm{FC}$  to $3\times r_\mathrm{FC}$, increases the simulated susceptibility by $<3\%$ and does not significantly affect the spatial structure~\cite{supp}. Second, we assume that only islands whose centers lie within a radius $r_\mathrm{cutoff}=0.9\,\mu\mathrm{m}$ of one another form junctions. We chose $r_\mathrm{cutoff}=0.9\,\mu\mathrm{m}$ so that $2\xi_N<r_\mathrm{cutoff}<2a$, where $\xi_N=\sqrt{\hbar D/k_\mathrm{B}T}\approx0.425\,\mu\mathrm{m}$ is the normal metal coherence length and $a=0.5\,\mu\mathrm{m}$ is the ordered array lattice constant~\cite{De_Gennes1964-la,Lobb1983-jf,eley_dependence_2013}. For this patch radius and junction cutoff radius, a typical patch contains a few thousand islands and $10,000-20,000$ junctions~\cite{supp}. Note that this model simulates the static magnetic response of the arrays, but the local field applied by the SQUID field coil varies sinusoidally at a frequency $f_\mathrm{FC}=800\,\mathrm{Hz}$. Thus, an additional assumption is that $f_\mathrm{FC}$ is slow compared to other timescales in the system, such that the applied field can be approximated as time-independent. This model also neglects any inductive coupling between the junctions, which we expect to be small given the very weak screening in this system~\cite{phillips_influence_1993}.

\begin{figure*}
    \includegraphics[width=\textwidth]{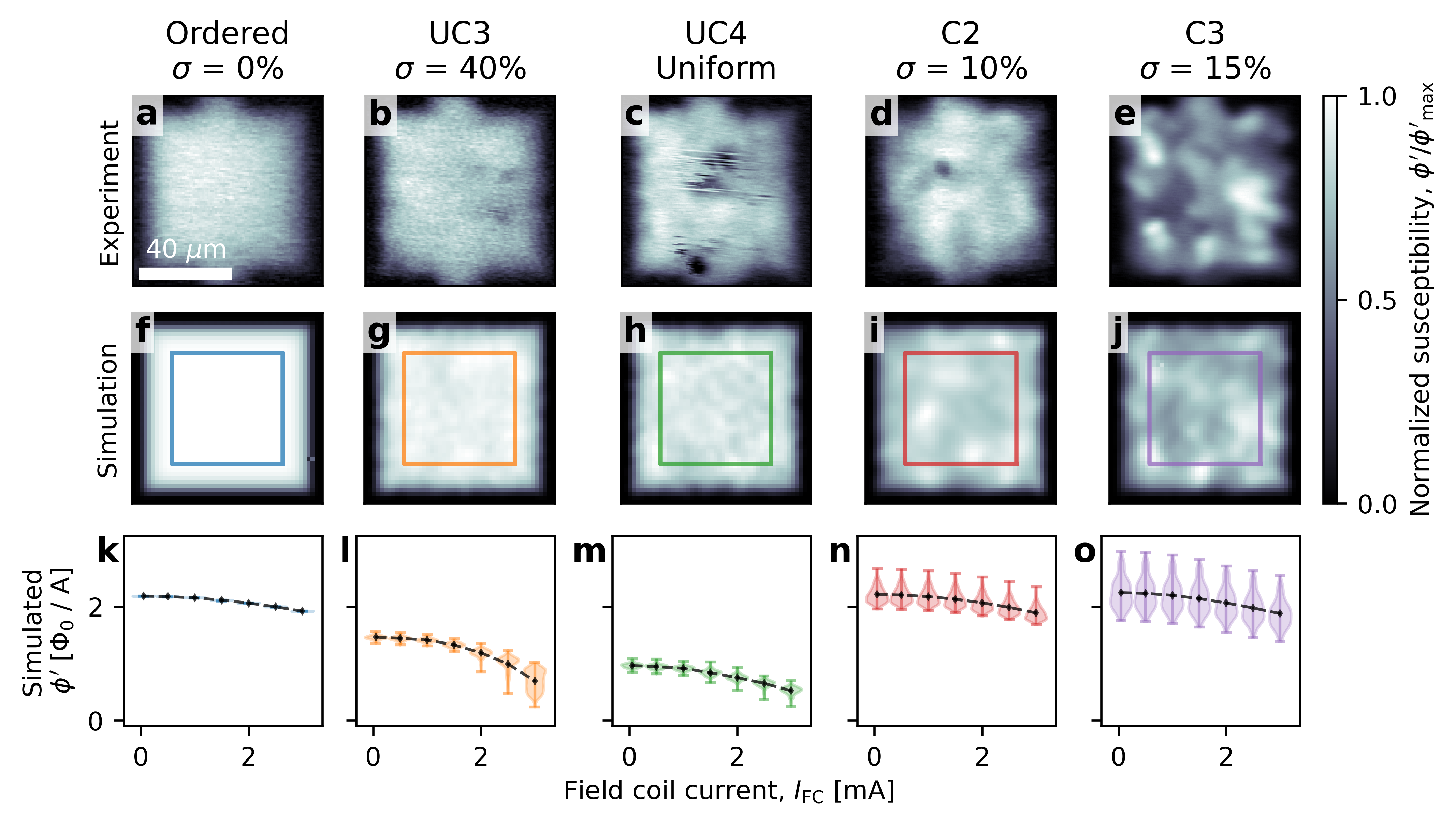}
    \caption{Comparison between the measured and simulated in-phase susceptibility $\phi'$ for all five arrays. Top row (a-e): Normalized in-phase susceptibility $\phi'/\phi'_\mathrm{max}$ measured at $I_\mathrm{FC}=0.05$ mA for panels (a, b, c, e) and $I_\mathrm{FC}=0.1$ mA for panel d. Middle row (f-j): $\phi'/\phi'_\mathrm{max}$ calculated using the junction network model with a field coil current of $I_\mathrm{FC}=0.05$ mA, a distance of $z_\mathrm{FC}=2\,\mu\mathrm{m}$ between the sample surface and SQUID field coil, and a critical current scale of $I_0=260\,\mathrm{nA}$. Bottom row (k-o): Distribution of simulated in-phase susceptibility $\phi'$ vs. field coil current, demonstrating the nonlinearity of the junction network model, the onset of which occurs at a higher field coil current than in experiment (compare to Figure~\ref{fgr:SSMdists}, where $\phi'$ is heavily suppressed in the uncorrelated arrays (b, c) well below $I_\mathrm{FC}=2\,\mathrm{mA}$). As in Figure~\ref{fgr:SSMdists}, we plot only the distribution of $\phi'$ in the central $50\times50\,\mu\mathrm{m}^2$ region of each array (see colored boxes in (f-j)). The black diamonds and dashed lines indicate the average susceptibility in these regions. Note that there are several regions near the center of array UC4 (panel c) where niobium islands have been scraped off due to contact with the SQUID susceptometer, which is not accounted for in the model. See Figure~\ref{fgr:sim-setup} and Supplemental Material~\cite{supp} for more details on the simulations.}
    \label{fgr:SSMsim}
\end{figure*}

A comparison of the in-phase susceptibility $\phi'$ measured in the SSM experiments and simulated using the junction network model is shown in Figure~\ref{fgr:SSMsim}. The model reproduces the spatially uniform and linear magnetic response in the ordered array (Figure~\ref{fgr:SSMsim} (a, f, k)). In the arrays with uncorrelated disorder, the local diamagnetic response exhibits a granular spatial structure (Figure~\ref{fgr:SSMsim} (b, c)), and $\phi'$ is suppressed more rapidly with increasing applied field than in the ordered array or the arrays with correlated disorder (Figure~\ref{fgr:SSMdists} (b, c)). Both of these effects are captured qualitatively by the junction network model (Figure~\ref{fgr:SSMsim} (g, h) and (l, m)). The simulated susceptibility of the correlated arrays is in good agreement with the measurements (Figure~\ref{fgr:SSMsim} (d, i) and (e, j)), which is most easily visible in array C3 (Figure~\ref{fgr:SSMsim} (e, j)) as it exhibits the highest contrast and most distinctive spatial features in $\phi'$ as a function of field coil position.

The magnitude of $\phi'$ depends strongly on the distance $z_\mathrm{FC}$ between the field coil and the sample, and in this particular experiment there was significant uncertainty ($\sim2-3\,\mu\mathrm{m}$) in $z_\mathrm{FC}$. This is likely the cause of the discrepancy in the magnitude of $\phi'$ at low field coil currents between Figure~\ref{fgr:SSMdists} (experiment) and Figure~\ref{fgr:SSMsim} (k-o) (simulation), and is why we plot the normalized susceptibility $\phi'/\phi'_\mathrm{max}$ in Figure~\ref{fgr:SSMsim} (a-j). We expect that if $z_\mathrm{FC}$ were known with a high degree of certainty, the junction network model would reproduce quantitatively both the magnitude and spatial structure of the in-phase susceptibility at low applied fields.

Although susceptibility simulations using the junction network model do exhibit nonlinearity with increasing applied field as the gauge-invariant phase $\theta$ across junctions in the network becomes large enough that $\sin\theta\approx\theta$ is not a good approximation (see Figure~\ref{fgr:SSMsim} (k-o) and Supplemental Material~\cite{supp}), in all cases the observed onset of nonlinearity in simulation occurs at a larger applied field than in experiment. For example, the ordered array shows appreciable out-of-phase response at $I_\mathrm{FC}=1.512\,\mathrm{mA}$ (Figure~\ref{fgr:SSMdata} (f)), however the flux through an $a\times a=500\times500\,\mathrm{nm}^2$ square unit cell or ``plaquette'' due to a loop with radius $r_\mathrm{FC}=6.8\,\mu\mathrm{m}$ carrying $I_\mathrm{FC}=1.512\,\mathrm{mA}$ is $<0.017\,\Phi_0$, corresponding to a gauge-invariant phase difference of $\theta<2\pi\times0.017/4\approx0.027$ radians across each of the four junctions in the plaquette, a value for which $\sin\theta\approx\theta$ is a very good approximation. 
Furthermore, this mechanism for nonlinearity is purely geometric, having no dependence on the overall strength of the Josephson coupling in the system (which is set in the model by the critical current scale $I_0$).

This suggests that there is another mechanism contributing to the onset of nonlinearity and dissipation in these arrays. One candidate is motion of vortices induced in the 2D superconducting system above a lower critical field $B^\mathrm{crit}_0$, as has been studied theoretically for uniform applied fields~\cite{fetter1980-kt, Schweigert1999-wu} and for nonuniform applied fields in the context of two-coil mutual inductance susceptibility measurements~\cite{Lemberger2013-lu, Lemberger2013-ha, Draskovic2013-ev}. Lemberger and Ahmed~\cite{Lemberger2013-ha} found that for a 2D superconductor in the weak screening (large $\Lambda$) limit with Ginzburg-Landau coherence length $\xi$ subject to a nonuniform field from a point dipole or small current loop, there can be no vortex-free state above a lower critical field $B^\mathrm{crit}_0\approx\Phi_0/(2\rho_0\xi)$, where $\rho_0\ll\Lambda$ is the radial distance from the magnetic source at which the applied field changes sign. This relationship between the coherence length $\xi$ and the onset of vortex-related nonlinear magnetic response has been used to measure $\xi$ in superconducting thin films~\cite{Draskovic2013-ev}. For our field coil with radius $r_\mathrm{FC}=6.8\,\mu\mathrm{m}$ located $2\,\mu\mathrm{m}$ above the film, the applied field changes sign at $\rho_0\approx7.5\,\mu\mathrm{m}$. Assuming $\xi=a$ for the ordered square array with lattice constant $a$~\cite{Lobb1983-jf} gives a lower critical field $B^\mathrm{crit}_0\approx 275\,\mu\mathrm{T}$, or a lower critical field coil current $I^\mathrm{crit}_\mathrm{FC}\approx3.4\,\mathrm{mA}$. The actual applied field (or field coil current) at which nonlinearity due to vortex dynamics begins to occur is necessarily smaller than this maximum vortex-free field, which is consistent with the SQUID measurements (Figure~\ref{fgr:SSMdata} (f)). In a homogeneous 2D superconductor, vortices are expected to first appear at the position with the highest superfluid momentum, i.e., the position where the vector potential $|\mathbf{A}(\mathbf{r})|$ is largest~\cite{tinkham2004introduction,Lemberger2013-ha}, and we expect the same to be true for the ordered array. Any vortices that are present will experience a force due to the local AC applied field from the field coil. If the vortices are not strongly pinned they will move under this force, and it is this vortex motion that causes dissipation. For a detailed analysis of the impact of vortex dynamics on two-coil mutual inductance measurements of 2D superconductors, see Ref.~\cite{Lemberger2016-pb}. Viewed through this lens, our results (Figures~\ref{fgr:SSMdata} and \ref{fgr:SSMdists}) suggest that disorder in the island spacing affects both the superfluid density and the coherence length $\xi$ of the 2D superconductor formed by the proximity coupled nano-islands, and that both quantities can be probed locally with SSM via the linear and nonlinear magnetic response, respectively.

Future work will focus on the magnetic response of ordered proximity coupled island arrays as a function of the island spacing over a wider range of local applied AC field. Such measurements will allow us to validate our model for the island spacing dependence of the junction critical currents, $I_c(d)$, and quantify how well the junction network model describes nonlinearities at higher applied field, which will in turn inform the design of future experiments on arrays with different geometries and materials. Beyond scanning SQUID microscopy, the proximity effect model systems introduced here could potentially be studied with time or frequency resolved imaging to better understand the dynamics causing dissipative behavior~\cite{Galin2020-qd}.

In summary, we have demonstrated the design, control, and measurement of a model superconducting system with engineered disorder to simulate the spatial evolution of superfluid density and phase coherence in 2D superconductors with micron-scale disorder. Scanning SQUID microscopy measurements reveal a magnetic response that is nonlinear and spatially inhomogeneous, and this response can be tuned by changing the disorder landscape. 
For small applied fields, the local diamagnetic response of the arrays is in good agreement with a model that treats the system as a network of Josephson junctions with island spacing-dependent Josephson coupling. However, we find that the onset of nonlinearity and dissipation with increasing applied field cannot be fully explained simply by considering the nonlinear (i.e., sinusoidal) current-phase relation of the junctions. This motivates future work on a theoretical description that incorporates the rich nonlinear and dissipative physics underlying this engineered, disordered 2D superconducting system.

\begin{acknowledgments}
The fabrication of this project was carried out in part in the Material Research Laboratory Central Research Facilities, University of Illinois. Some of the computing for this project was performed on the Sherlock cluster. We would like to thank Stanford University and the Stanford Research Computing Center for providing computational resources and support that contributed to these research results.

This work was primarily supported by the DOE ``Quantum Sensing and Quantum Materials'' Energy Frontier Research Center under Grant DE-SC0021238. Samples were fabricated with support from DOE Basic Energy Sciences under DE-SC0012649. The SSM measurements were, in part, supported by the Gordon and Betty Moore Foundation, Grant GBMF3429 ``Exotic Emergent Particles in Nanostructures''. Work at the University of Connecticut was supported by the State of Connecticut. 

\end{acknowledgments}

%
%

\bibliography{NbonAuPaperSources}

\newpage
\part*{Supplemental Material}
\setcounter{section}{0}
\setcounter{figure}{0}
\setcounter{equation}{0}
\setcounter{page}{1}
\renewcommand{\thesection}{S\arabic{section}}
\renewcommand{\thetable}{S\arabic{table}}
\renewcommand{\thefigure}{S\arabic{figure}}

\title{Supplemental Material for ``Local imaging of diamagnetism in proximity coupled niobium nano-island arrays on gold thin films''}

\author{Logan Bishop-Van Horn}
\altaffiliation{Contributed equally to this work}
\affiliation{Department of Physics, Stanford University, Stanford, California 94305, USA}
\affiliation{Stanford Institute for Materials and Energy Sciences, SLAC National Accelerator Laboratory, Menlo Park, California 94025, USA}

\author{Irene P. Zhang}
\altaffiliation{Contributed equally to this work}
\affiliation{Department of Applied Physics, Stanford University, Stanford, California 94305, USA}
\affiliation{Stanford Institute for Materials and Energy Sciences, SLAC National Accelerator Laboratory, Menlo Park, California 94025, USA}

\author{Emily N. Waite}
\affiliation{Department of Physics and Materials Research Laboratory, University of Illinois, Urbana, Illinois 61801, USA}

\author{Ian Mondragon-Shem}
\affiliation{Department of Physics and Astronomy, Northwestern University, Evanston, Illinois 60208, USA}
\affiliation{Materials Science Division, Argonne National Laboratory, Argonne, Illinois 60439, USA}

\author{Scott Jensen}
\affiliation{Department of Physics and Materials Research Laboratory, University of Illinois, Urbana, Illinois 61801, USA}

\author{Junseok Oh}
\affiliation{Department of Physics and Materials Research Laboratory, University of Illinois, Urbana, Illinois 61801, USA}

\author{Tom Lippman}
\affiliation{Department of Physics, Stanford University, Stanford, California 94305, USA}

\author{Malcolm Durkin}
\affiliation{Department of Physics, University of Colorado Boulder, Boulder, Colorado 80305, USA}

\author{Taylor L. Hughes}
\affiliation{Department of Physics and Materials Research Laboratory, University of Illinois, Urbana, Illinois 61801, USA}

\author{Nadya Mason}
\affiliation{Department of Physics and Materials Research Laboratory, University of Illinois, Urbana, Illinois 61801, USA}

\author{Kathryn A. Moler}
\affiliation{Department of Applied Physics, Stanford University, Stanford, California 94305, USA}
\affiliation{Stanford Institute for Materials and Energy Sciences, SLAC National Accelerator Laboratory, Menlo Park, California 94025, USA}

\author{Ilya Sochnikov}
\email{ilya.sochnikov@uconn.edu}
\affiliation{Department of Physics, University of Connecticut, Storrs, Connecticut 06269, USA}
\affiliation{Institute of Material Science, University of Connecticut, Storrs, Connecticut 06269, USA}

\date{\today}

\maketitle


\section{Scanning SQUID susceptometry}

In scanning Superconducting Quantum Interference Device (SQUID) susceptometry, a small single-turn field coil (FC) locally applies a magnetic field to a sample, and a pickup loop (PL)---concentric with the field coil and connected to flux-sensitive SQUID circuit---senses the sample's magnetic response~\cite{Gardner2001-gr}. The SQUID susceptometer used in this study~\cite{huber2008gradiometric} is gradiometric, with two counter-wound field coil-pickup loop pairs separated by $\sim$ 1 mm, only one of which (the ``front field coil'') is brought close to the sample surface. Each FC-PL pair has a mutual inductance $|\Phi_\mathrm{PL}/I_\mathrm{FC}|$ of approximately $900\,\Phi_0/\mathrm{A}$, meaning that a current of $I_\mathrm{FC}=1\,\mathrm{mA}$ flowing through the field coil threads a flux of $|\Phi_\mathrm{PL}|=0.9\,\Phi_0\approx1.86\,\mathrm{mT}\,\mu\mathrm{m}^2$ through the pickup loop. The gradiometric geometry of the circuit means that, in the absence of a sample with nonzero magnetic susceptibility, the total mutual inductance between the two field coils and the SQUID is zero (modulo any lithographic imperfections): $M = (\Phi^\mathrm{front}_\mathrm{PL} + \Phi^\mathrm{back}_\mathrm{PL})/I_\mathrm{FC}=0\,\Phi_0/\mathrm{A}$.

When the front field coil is brought close to a superconducting sample, the sample screens the field from the field coil, modifying the mutual inductance of the front FC-PL pair, and thus the mutual total mutual inductance of the susceptometer. The amount by which the sample modifies the SQUID mutual inductance $M$ is the scanning SQUID susceptibility signal $\phi(x, y) = \Delta M_{z_\mathrm{fC}}(x, y)$, which is measured as a function of relative sample-sensor position $(x, y)$ as the sensor is rastered over the sample surface with the field coil at a fixed height $z_\mathrm{FC}$. $\phi(x, y)$ is measured using phase-sensitive low-frequency lock-in detection, allowing us to detect both the in-phase magnetic response $\phi'(x, y)$ and out-of-phase magnetic response $\phi''(x, y)$ of the sample. In addition to the sample-sensor spacing $z_\mathrm{FC}$, the susceptibility signal $\phi$ depends on the sample's local magnetic screening length (the London penetration depth $\lambda$ for bulk superconductors, or the Pearl length $2\Lambda=2\lambda^2/t$ for superconducting thin films with thickness $t\ll\lambda$)~\cite{Kirtley2012-od}. Given the layer structure of the SQUID susceptometer used here, the pickup loop is roughly 550 nm closer to the sample than the field coil: $z_\mathrm{PL}=z_\mathrm{FC}-550\,\mathrm{nm}$, which is accounted for in the modeling~\cite{huber2008gradiometric}.

\section{Additional data}
\label{sec:supp-data}

\begin{figure}
  \includegraphics[width=\linewidth]{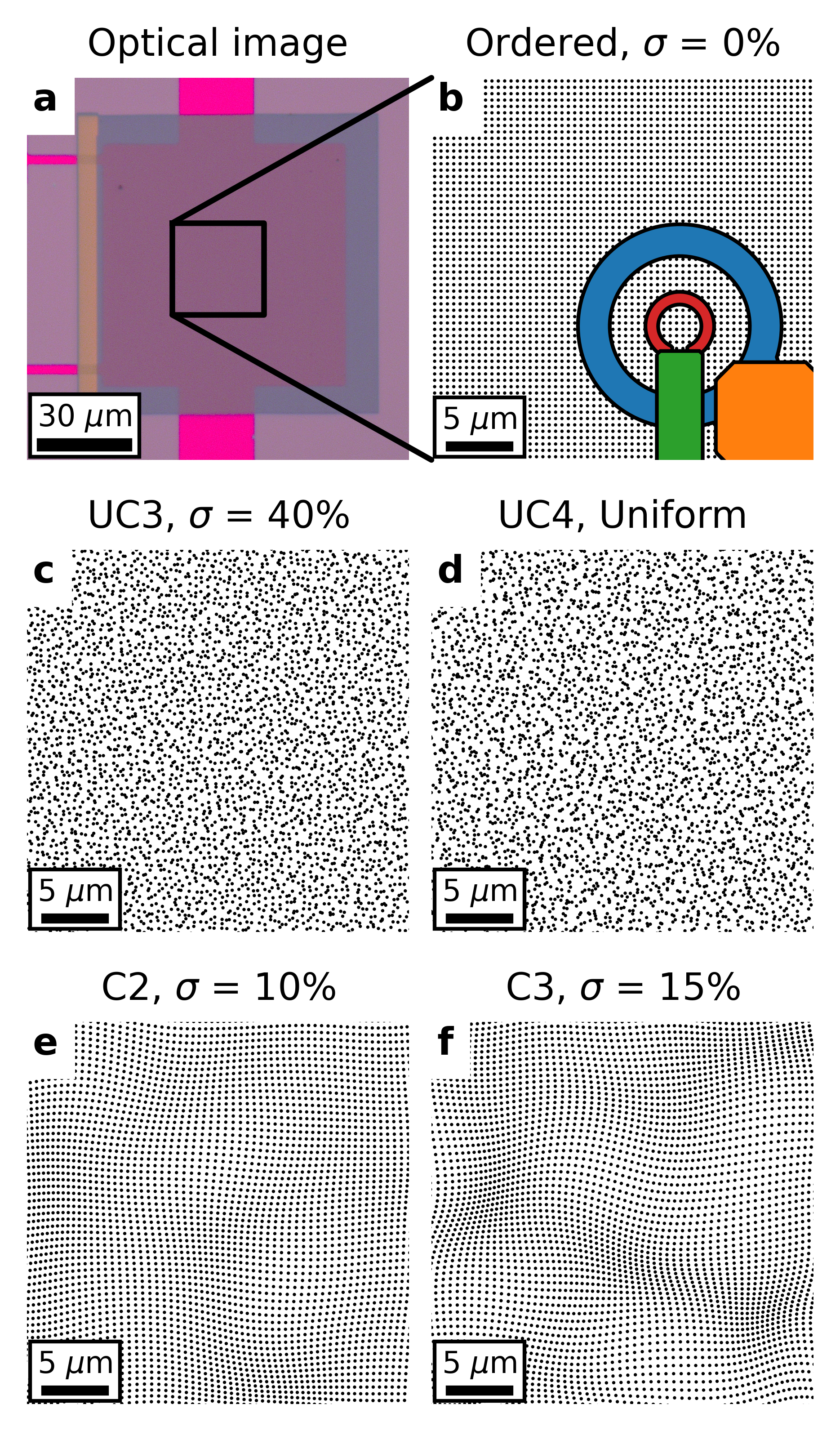}
  \caption{Device and sensor geometry. (a) Optical image of the ordered array. The background is the SiO${}_{2}$ substrate, the bright pink regions are bare gold film, the dark gray region is niobium islands directly on SiO${}_{2}$, and the dark pink central $80\,\mu\mathrm{m}\times80\,\mu\mathrm{m}$ region is niobium islands on top of the gold film. (b-f) Schematic of the designed island positions for a $30\,\mu\mathrm{m}\times 30\,\mu\mathrm{m}$ region of (b) the ordered array, (c) the 40\% standard deviation uncorrelated array UC3, (d) the uniformly-distributed array UC4, (e) the 10\% standard deviation correlated array C2, and (f) the 15\% standard deviation correlated array C3. The islands are drawn to scale, i.e. each island has a diameter of 260 nm. The SQUID susceptometer geometry is shown in (b), with the field coil in blue (larger radius) and the pickup loop in red (small radius). There are also superconducting shields covering the field coil and pickup loop leads, which are shown as orange and green polygons respectively. Note the difference in lateral scale between Figure~\ref{fgr:islands-squid} (b-f) and Figures~\ref{fgr:SSMdata-in-phase} and \ref{fgr:SSMdata-out-of-phase} below.
}
\label{fgr:islands-squid}
\end{figure}

Figure~\ref{fgr:islands-squid} shows an expanded version of Figure 1 in the main text, including niobium island positions for all five arrays listed in Table 1 of the main text.

\subsection{SEM images}

\begin{figure*}
    \centering
    \includegraphics[width=0.8\linewidth]{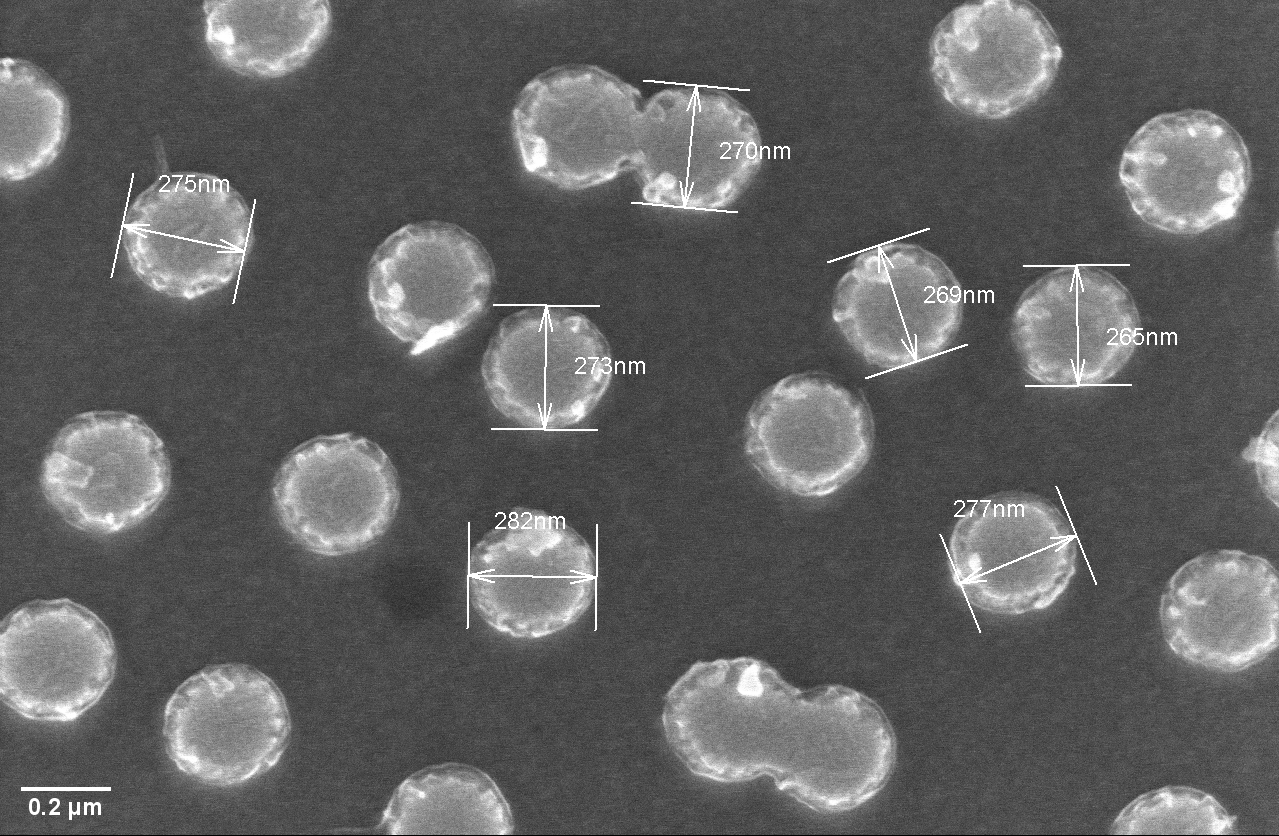}
    \caption{Scanning electron microscopy (SEM) image taken of device C2 (10\% correlated disorder) showing roughly 20 islands. Measurements of the diameter of the islands were taken to show the consistency of the size of the islands. All islands measured have a diameter between 265 nm and 282 nm at their widest point.}
    \label{fig:semZoomin}
\end{figure*}
\begin{figure*}
    \centering
    \includegraphics[width=0.8\textwidth]{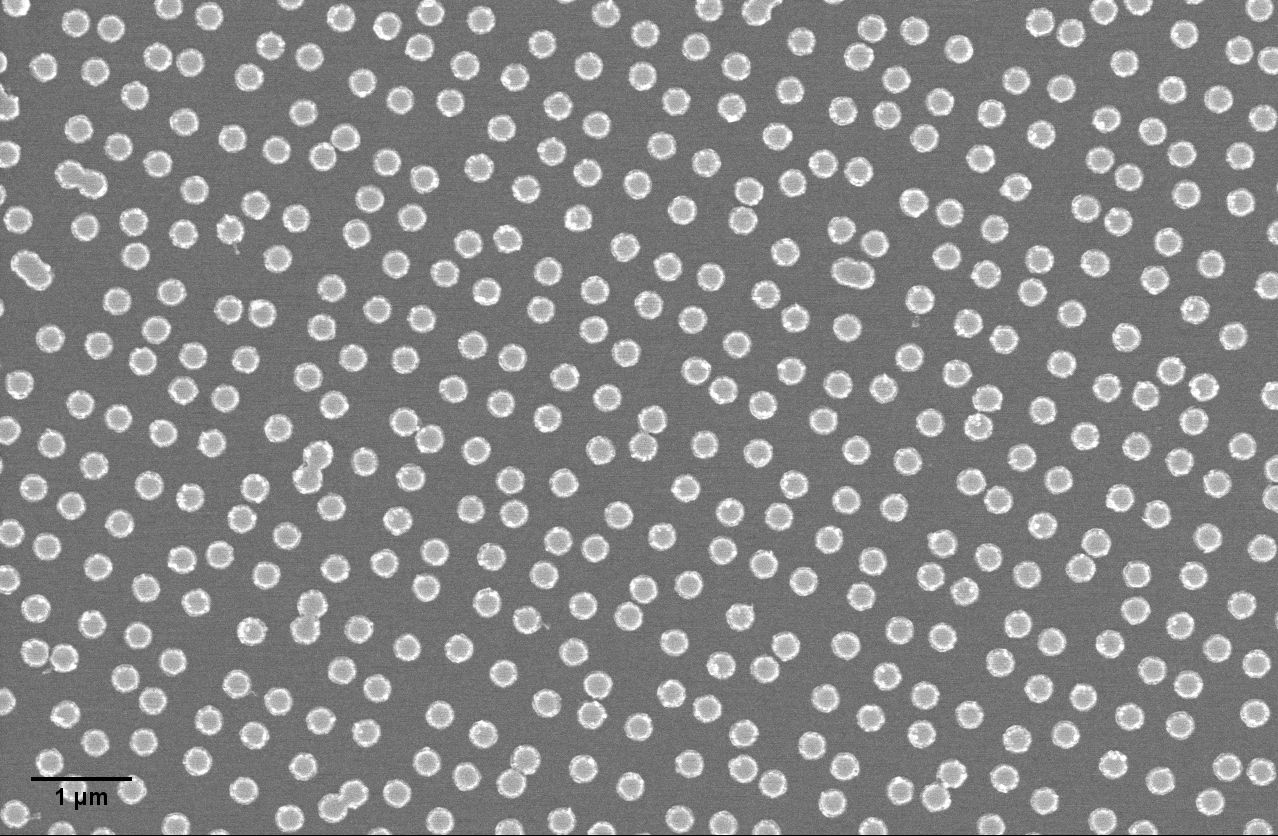}
    \caption{SEM image of device C2 showing a roughly 12 $\mu$m by 8 $\mu$m area with hundreds of islands.}
    \label{fig:semZoomOut}
\end{figure*}

Figure \ref{fig:semZoomin} shows a scanning electron microscopy (SEM) image of the islands at a magnification of approximately $\times$45,000. This image was taken on a region of C2 where the gold substrate underneath is present. On top of the image are measurements placed to show the spread of the island size. The difference between the largest and smallest islands measured is 17 nm, indicating that while there is some variance in island size, the variance is negligible compared to other length scales involved. Figure \ref{fig:semZoomOut} shows an SEM image taken at $\times$10,000 magnification on C2. One feature of note is that islands with center-to-center spacing smaller than their average diameter will merge together to form one larger non-cylindrical island, which will behave differently than two individual smaller islands \cite{durkin_2020}. 

\subsection{SSM measurements}

\begin{figure*}
  \includegraphics[width=\linewidth]{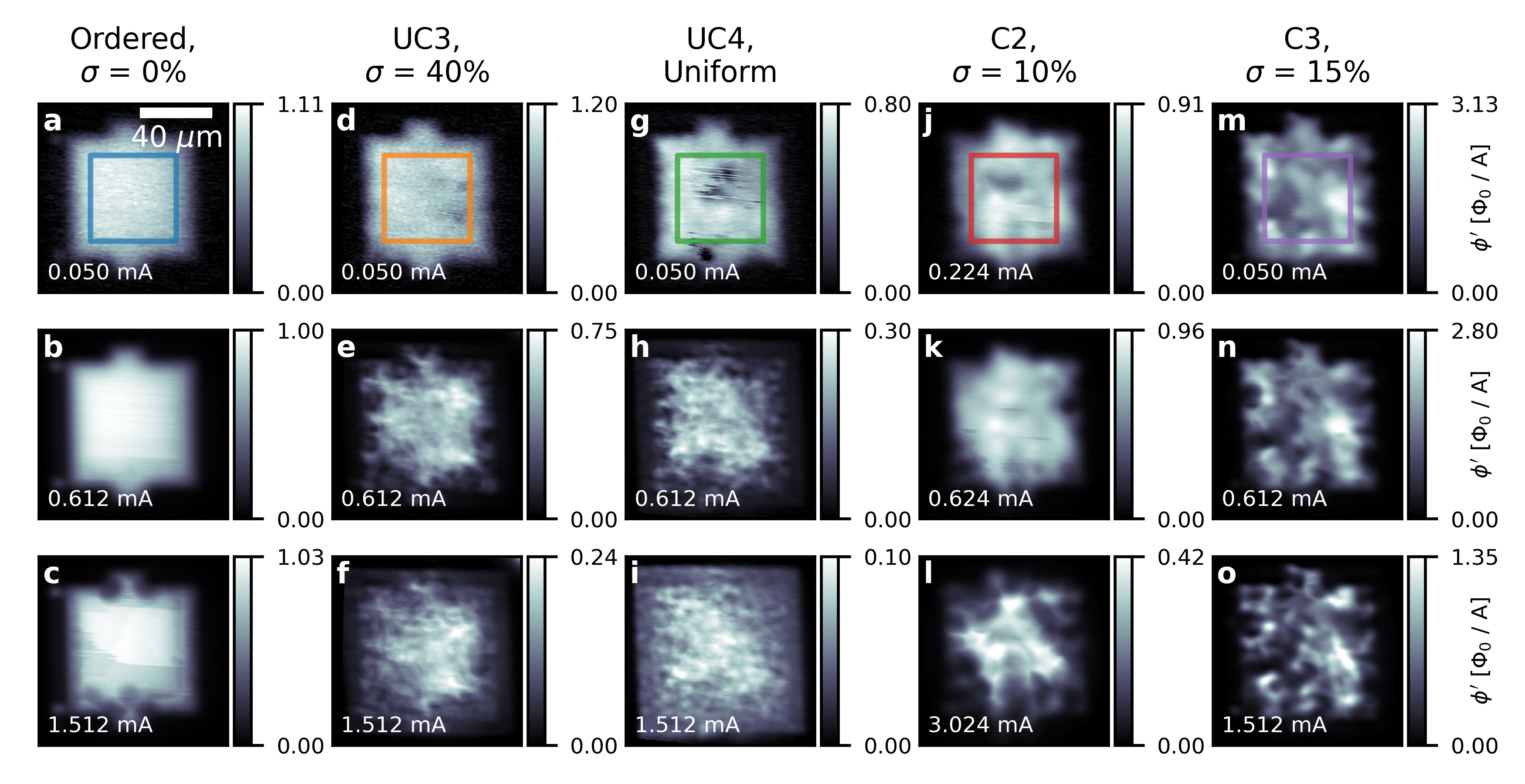}
  \caption{In-phase susceptibility $\phi'$ for all five samples listed in Table 1 in the main text for three selected RMS field coil currents $I_\mathrm{FC}$. All images were taken at the same nominal sensor height and temperature, except for C2 (j, k, l) as explained in the main text of the Supplemental Material. The colored boxes in the first row (a, d, g, j, m) indicate the $50\times50\,\mu\mathrm{m}^2$ regions for which the distributions of $\phi'$ and $\phi''$ are shown as a function of $I_\mathrm{FC}$ in Figure~\ref{fgr:SSMdists-in-out}.
}
\label{fgr:SSMdata-in-phase}
\end{figure*}

\begin{figure*}
  \includegraphics[width=\linewidth]{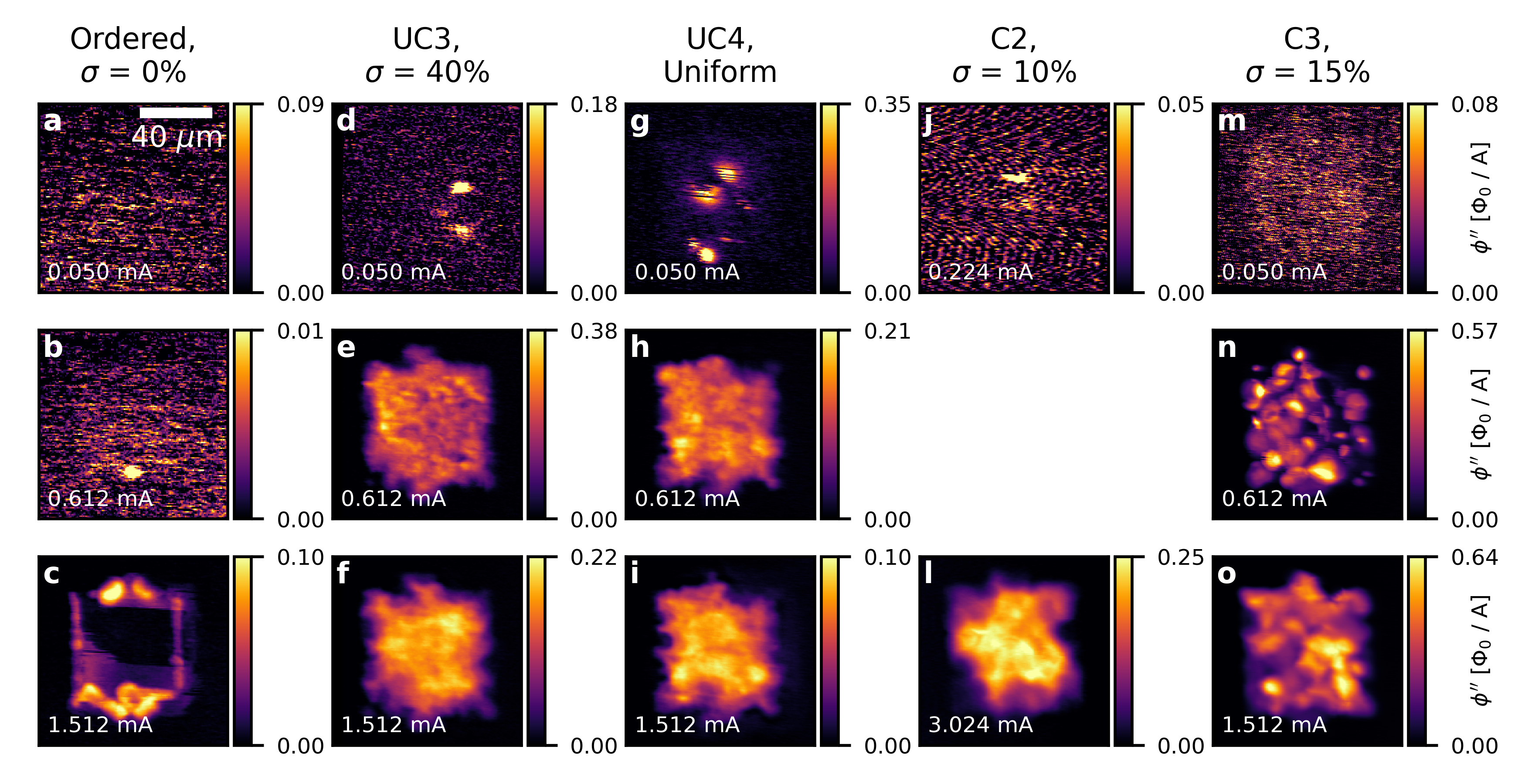}
  \caption{Out-of-phase susceptibility $\phi''$ for all five samples listed in Table 1 in the main text for three selected RMS field coil currents $I_\mathrm{FC}$. All images were taken at the same nominal sensor height and temperature, except for C2 (j, l) as explained in the main text of the Supplemental Material. $\phi''$ was only recorded at two field coil currents for C2 (j, l). Note that panels (a, b, d, j, m) only show noise, as $\phi''$ is below the measurement noise floor for these small values of $I_\mathrm{FC}$.
}
\label{fgr:SSMdata-out-of-phase}
\end{figure*}

Figures~\ref{fgr:SSMdata-in-phase} and \ref{fgr:SSMdata-out-of-phase} show the in-phase susceptibility $\phi'$ and out-of-phase susceptibility $\phi''$ respectively for all five samples listed in Table 1 of the main text for three selected field coil currents (indicated in the lower left of each image). All images were taken at the same nominal temperature and SQUID - sample distance $z_\mathrm{FC}$, with the exception of the data for C2 (panels j, k, l), which is known to have been measured with a $2-3\,\mu\mathrm{m}$ larger sensor-sample distance compared to all other arrays, resulting in a relative overall reduction in signal magnitude. However, for all measurements shown here there is some uncertainty ($\sim2-3\,\mu\mathrm{m}$) in $z_\mathrm{FC}$, which makes it difficult to compare the absolute magnetic of the diamagnetic response both between measurements of different arrays and between measurements and simulations of the same array. Note that the field coil currents shown for C2 (j, k, l) differ slightly from those shown for the other four samples.

The signal-to-noise ratio (SNR) of a SQUID susceptibility measurement increases with increasing field coil current $I_\mathrm{FC}$. At large field coil currents (Figure~\ref{fgr:SSMdata-in-phase}, bottom row), the diamagnetic response of the bare niobium islands deposited outside the region with gold film is visible in the images of the two uncorrelated arrays, (UC3, Figure~\ref{fgr:SSMdata-in-phase} (f)) and (UC4, Figure~\ref{fgr:SSMdata-in-phase} (i)). This is why the region of nonzero $\phi'$ is larger in (f, i) than it is in (d, g). The dissipative out-of-phase response $\phi''$ is limited to the smaller region with gold beneath the islands (Figure~\ref{fgr:SSMdata-out-of-phase}f and i). For the ordered and correlated disorder samples (c, l, o), the signal from the bare niobium islands is not easily visible due to the large signal from the proximitized gold region.
\begin{figure*}
  \includegraphics[width=0.95\linewidth]{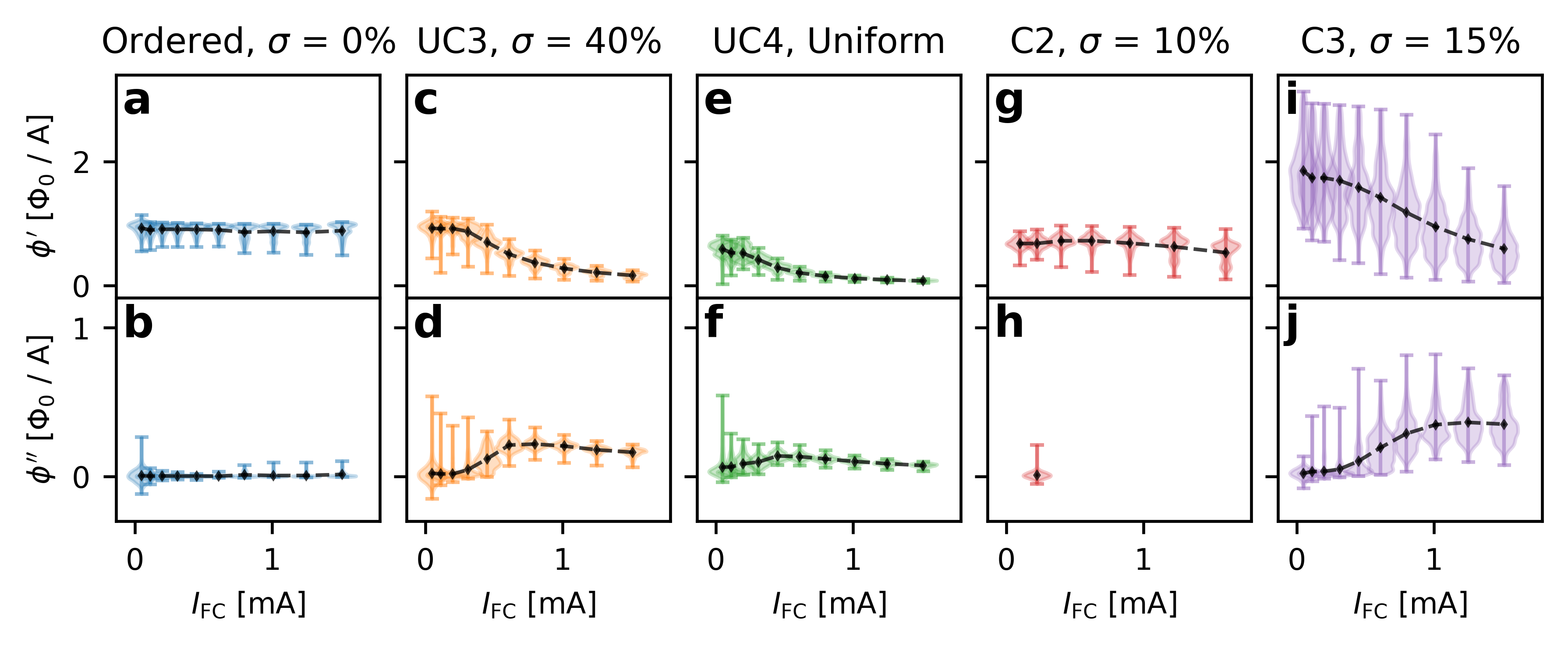}
  \caption{Violin plots showing the measured distribution of both in-phase susceptibility $\phi'$ and out-of-phase susceptibility $\phi''$ as a function of applied field coil current $I_\mathrm{FC}$ for the central $50\times50\,\mu\mathrm{m}$ region of all five arrays (see colored boxes in the top row of Figure~\ref{fgr:SSMdata-in-phase}). The apparent negative out-of-phase susceptibility at low field coil currents is due to the poor SNR of the susceptometry for small $I_\mathrm{FC}$. The black diamonds and dashed lines shown the mean of each distribution.}
\label{fgr:SSMdists-in-out}
\end{figure*}
Figure~\ref{fgr:SSMdists-in-out} is an expanded version of Figure 3 in the main text, showing the distribution of both in-phase susceptibility $\phi'$ and out-of-phase susceptibility $\phi''$ as a function of applied field coil current for all five arrays. Note that $\phi''$, which is associated with dissipation in the arrays, starts near zero at low field coil current and appears to plateau at higher field coil current. In contrast, $\phi'$, which is associated with Meissner screening currents, starts at a finite value at low field coil currents and decreases with increasing field coil current (except near the center of the ordered array, where $\phi'$ is roughly constant as a function of field coil current).

As mentioned in the main text, the ordered array begins to exhibit an inhomogeneous magnetic response at the highest local applied AC field (or field coil current $I_\mathrm{FC}$), as can be seen in Figure 2 (c, f) or in panel (c) of Figures~\ref{fgr:SSMdata-in-phase} and \ref{fgr:SSMdata-out-of-phase}. The edges of the array show significantly reduced in-phase response $\phi'$  (Figure~\ref{fgr:SSMdata-in-phase} (c)) and increased dissipative out-of-phase response $\phi''$ (Figure~\ref{fgr:SSMdata-out-of-phase} (c)). Even away from the edges, the bottom half of the array shows a slightly smaller in-phase response than the top half of the array (Figure~\ref{fgr:SSMdata-in-phase} (c)).

This difference in $\phi'$ between the top and bottom halves of the ordered array is most likely due to local heating of the gold film caused by vortex motion near the edges of the array, rather than to an intrinsic difference between the two halves of the device. If the junctions in the bottom half of the array simply had a smaller critical current (e.g., due to lithography or morphology differences), the bottom half of the array would have a weaker diamagnetic response than the top half for all values of the applied AC field (as the effective penetration depth for an ordered junction array with junction critical currents $I_c$ is $\Lambda\propto 1/I_c$). However, the diamagnetic response of the ordered array appears to be very homogeneous for smaller applied AC fields (Figure~\ref{fgr:SSMdata-in-phase} (a, b)).

To generate the susceptibility maps shown in Figures~\ref{fgr:SSMdata-in-phase} and \ref{fgr:SSMdata-out-of-phase}, the SQUID sensor is raster-scanned over the field of view. Typically starting at the bottom left corner of the field of view, data is acquired while the field coil and pickup loop are scanned horizontally to the bottom right corner, producing the bottom row of pixels. Then the sensor is returned to the left side of the field of view (without acquiring data) and its position is advanced towards the ``top'' of the field of view by a distance corresponding to one pixel. The process is repeated to acquire the next row of pixels. Vortex motion occurring when the sensor is near the bottom edge of the array (orange/yellow features at the bottom of Figure~\ref{fgr:SSMdata-out-of-phase} (c)) could heat the gold film causing a weaker diamagnetic response from the bottom half of the array, but only at applied AC fields large enough to induce vortices near the edge of the array.

The diagonal feature near the center right of the array in panel (c) of Figures~\ref{fgr:SSMdata-in-phase} and \ref{fgr:SSMdata-out-of-phase} is also consistent with this picture (i.e., heating due to vortex motion as the sensor passes over the left edge of the array). Under this hypothesis, the fact that $\phi'$ in the top right quadrant of the array in Figure~\ref{fgr:SSMdata-in-phase} (c) is equal to $\phi'$ in the entire array at lower applied AC fields in Figures~\ref{fgr:SSMdata-in-phase} (a, b) indicates that the heating effect is absent when the field coil and pickup loop are in that region of the device, i.e., the gold film has cooled to its equilibrium temperature by the time the SQUID reaches that region, and that region is measured before any heating due to vortex motion at the top edge of the array occurs. The SQUID sensor is scanned in a plane that is nominally parallel to the sample surface. A scan plane that is not exactly parallel to the sample surface, such that the SQUID sensor is slightly closer to the surface in some parts of the sample than in others, could also contribute to spatially nonuniform dissipation and heating even in a homogeneous sample.

\subsection{Transport measurements}

\begin{figure}
    \centering
    \includegraphics[width=\linewidth]{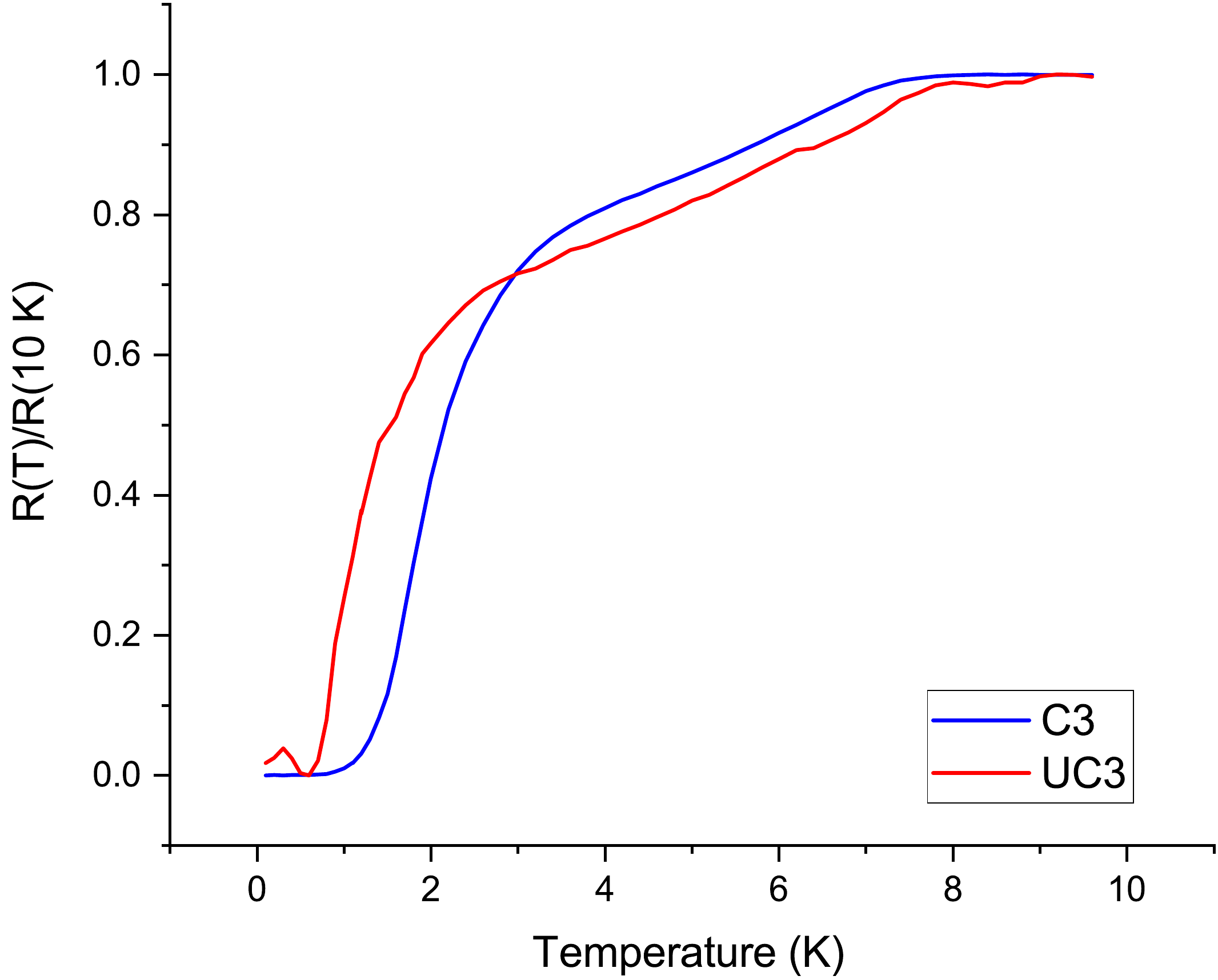}
    \caption{Resistance versus temperature data for device C3 and UC3. In C3, we can see that the islands begin to superconduct slightly below 8 K. This starts to level off around 5 K. Near 3.5 K, the slope increases from the gold experience a proximity effect from the islands until the sample reaches zero resistivity at around 0.8 K.  In UC3, the island superconducting regime is more smeared out and the proximity-induced effect of gold requires lower temperatures to begin. Note that the data for UC3 was taken using a 3-point measurement, so the data was offset by a constant resistance and had a phase shift near 0 Kelvin, resulting in a seemingly non-zero resistance.} 
    \label{fig:transport}
\end{figure}

Lastly, Figure \ref{fig:transport} shows transport data taken on two of the devices: C3 (correlated disorder with 15\% standard deviation) and UC3 (uncorrelated disorder with 40\% standard deviation). This data was taken in a dilution refrigeration system with a base temperature of 10 mK. Note that the data taken on UC3 was done using a 3-point measurement set-up, while C3 was done using a 4-point measurement set-up. As a result, the data from UC3 has a lower signal-to-noise ratio and the resistance was offset by a constant. Another result of the 3-point measurement is the resistance seemingly increases below 1 K. This is believed to be due to a change of phase in the current, and does not have any physical significance. 

It is important to note how the change in disorder affects the shape of the resistance versus temperature graph. In C3, the islands begin to enter the superconducting phase at around 8 K. The sharper drop around 3.5 K represents when the system was sufficiently cool enough to have a proximity effect on the surrounding gold to get a supercurrent flowing through parts of the gold substrate. In the uncorrelated sample, the superconducting transition of the islands has a larger range and it takes a lower temperature for the gold to become a proximity-induced superconductor. For further details on the resistance versus temperature graphs for superconducting island arrays, see Eley \textit{et al.}~\cite{eley_approaching_2012}.

\section{Junction network model}

\subsection{Model}

In the model discussed in the main text, which we call the discrete junction network (DJN) model, we treat the system as a network or graph of $n$ circular superconducting islands (nodes), all with the same diameter, $D_\mathrm{island}=260\,\mathrm{nm}$, with center positions $\mathbf{r}_i$ and superconducting phases $\varphi_i\in[0,2\pi)$. We assume that pairs of islands $(i,j)$ form 1D Josephson junctions (edges) with critical current $I_{c,ij}$ or Josephson energy $E_{J,ij}=I_{c,ij}\Phi_0/2\pi$. To simplify the problem, we assume that only islands whose centers lie within a radius $r_\mathrm{cutoff}$ of one another form junctions, such that there are $m$ junctions in total. As described in the main text, we assume that the junction critical currents depend on junction length or edge-to-edge island spacing $d_{ij}=|\mathbf{r}_i-\mathbf{r}_j|-D_\mathrm{island}$ according to $I_{c,ij}(d_{ij})=I_0(d_0/d_{ij})^{2}$, where $d_0=240\,\mathrm{nm}$ is the minimum edge-to-edge island spacing in the ordered array and $I_0$ is a constant that corresponds to the maximum critical current per junction in the ordered array~\cite{eley_dependence_2013}.

We remove from the network any islands that have fewer than 2 neighbors within the cutoff radius $r_\mathrm{cutoff}$ because such islands are either completely isolated (0 neighbors) or can only satisfy the current conservation constraint (see below) if no current flows through the adjacent junction (1 neighbor). If two or more islands are overlapping (i.e., edge-to-edge island spacing $d_{ij}\leq 0$, see Figure~\ref{fig:semZoomin}), we remove the overlapping islands and replace them with a single island whose center is located at the average position of the centers of the original islands.

For an applied magnetic vector potential $\mathbf{A}(\mathbf{r})$, assuming all junctions have a sinusoidal current-phase relation (CPR), the current $I_{ij}$ flowing from island $i$ to island $j$ is $I_{ij}=I_{c, ij}\sin\theta_{ij}$, where
$\theta_{ij}=\varphi_j-\varphi_i-2\pi A_{ij}$ is the gauge-invariant phase difference and $A_{ij}=\Phi_0^{-1}\int_{\mathbf{r}_i}^{\mathbf{r}_j}\mathbf{A}(\mathbf{r})\cdot\mathrm{d}\mathbf{r}$. Given a network constructed as described above, the goal is to find island phases $\varphi_i$ that satisfy the following constraints:
\begin{itemize}
	\item Current conservation: for each island $i$, the sum of the currents entering $i$ is equal to the sum of the currents leaving $i$.
\item Phase single-valuedness: for each loop (closed path) $\ell$ in the network,
    \begin{widetext}
    \begin{align*}
        \sum_{i, j\text{ in loop }\ell}\left(\varphi_j-\varphi_i-\frac{2\pi}{\Phi_0}\int_{\mathbf{r}_i}^{\mathbf{r}_j}\mathbf{A}(\mathbf{r})\cdot\mathrm{d}\mathbf{r}\right)
            &=\sum_{i, j\text{ in loop }\ell}\left(\delta_{ij}-2\pi A_{ij}\right)\\
            &=\sum_{i, j\text{ in loop }\ell}\theta_{ij}\\
            &=2\pi(f_\ell-n_\ell),
    \end{align*}
    \end{widetext}
    where $f_\ell\in\mathbb{R}$ is the frustration index of the loop and $n_\ell\in\mathbb{Z}$ is the number of flux quanta enclosed in the loop~\cite{tinkham2004introduction}.
\end{itemize}
For a connected graph with $n$ nodes and $m$ edges, the number of \emph{basis cycles} is $m-n+1$~\cite{Paton1969-kj}. All loops in the graph can be formed from these basis cycles via symmetric differences, so this represents the minimum number of loops for which we need to enforce the phase single-valuedness constraint. We treat the total flux $\Phi_\ell=\Phi_0f_\ell$ through each loop $\ell$ as a single continuous variable. Therefore, the number of degrees of freedom in the problem is:
\begin{widetext}
\begin{align*}
    N_\mathrm{DOF}=&+n\text{ [$\varphi_i$ for each island $i$]}\\
    &+(m-n+1)\text{ [$\Phi_\ell$ for each basis cycle $\ell$]}\\
    &-n\text{ [current conservation constraint for each island $i$]}\\
    &-(m-n+1)\text{ [phase single-valuedness constraint for each basis cycle $\ell$]}\\
    &=0.
\end{align*}
\end{widetext}
We solve this nonlinear programming (NLP) problem using the GEKKO Python interface to the APMonitor optimization suite~\cite{Beal2018-dq,Hedengren2014-zu}. 

The field coil and pickup loop are approximated as 1D circular loops with radii $r_\mathrm{FC}=6.8\,\mu\mathrm{m}$ and $r_\mathrm{PL}=2.5\,\mu\mathrm{m}$ respectively (see Figure~\ref{fig:clf-setup} (c)). The magnetic vector potential $\mathbf{A}(\mathbf{r})$ from the field coil carrying current $I_\mathrm{FC}$ is given in terms of the complete elliptic integrals of the first and second kind, $K(m)$ and $E(m)$~\cite{Jackson1999-zb}:
\begin{align*}
    \mathbf{A}(r,\theta,\varphi)&=-\frac{\mu_0I_\mathrm{FC}a}{\pi m}\frac{(m-2)K(m)+2E(m)}{\sqrt{r^2+a^2+2ar\sin\theta}}\hat{\mathbf{\varphi}},\text{ where}\\
    a&=r_\mathrm{FC},\text{ and}\\
    m&=k^2\\
    &=\frac{4ar\sin\theta}{r^2+a^2+2ar\sin\theta}.
\end{align*}
Here the vector $\mathbf{r}=(r,\theta,\varphi)$ is given in spherical coordinates relative to the center of the field coil, with $r$ the radial distance from the field coil center, $\theta$ the polar angle, and $\varphi$ the azimuthal angle. For each field coil position, we model a ``patch" containing all islands within a radius $r_\mathrm{patch}$ of the field coil center, rather than the entire array in order to reduce the computational resources required to solve the model (see discussion of scaling below). We assume that junctions outside of $r_\mathrm{patch}$ are both weakly influenced by the field from the field coil and inefficient at coupling flux into the pickup loop such that they don't contribute significantly to the susceptibility signal (see Figure~\ref{fig:djn-patch}).

The magnetic field $\mathbf{B}$ produced by a given set of junction currents $I_{ij}$, is given by the Biot-Savart law:
\begin{align*}
    \mathbf{B}(\mathbf{r})&=\frac{\mu_0}{4\pi}\sum_{\text{junctions }ij}\frac{\mathbf{I}_{ij}\times(\mathbf{r}-\mathbf{r}_{ij})}{|\mathbf{r}-\mathbf{r}_{ij}|^3},\text{ where}\\
    \mathbf{r}_{ij}&=\frac{1}{2}(\mathbf{r}_i+\mathbf{r}_j)\text{ is the center of junction }ij,\text{ and}\\
    \mathbf{I}_{ij}&=I_{ij}(\mathbf{r}_j-\mathbf{r}_i).
\end{align*}
The resulting flux through the pickup loop $\Phi_\mathrm{PL}$ is found by integrating the $z$-component of $\mathbf{B}$ over the surface of the pickup loop. In practice, the pickup loop is discretized into a Delaunay mesh~\cite{Shewchuk1996-va} composed of a few thousand triangles $t$, each with area $A_t$ and centroid (center of mass) position $\mathbf{r}_t$, so that the flux through the pickup loop is:
\begin{align*}
    \Phi_\mathrm{PL}&=\int_\mathrm{PL}\mathbf{B}(\mathbf{r})\cdot\mathbf{\hat{z}}\,\mathrm{d}^2r\\
    &\approx\sum_{\text{triangles } t}A_t\mathbf{B}(\mathbf{r}_t)\cdot\mathbf{\hat{z}}.
\end{align*}
Finally the (in-phase) susceptibility is given by $\phi'=\Phi_\mathrm{PL}/I_\mathrm{FC}$.

\subsection{Nonlinearity}

\begin{figure*}
	\centering
	\includegraphics[width=\linewidth]{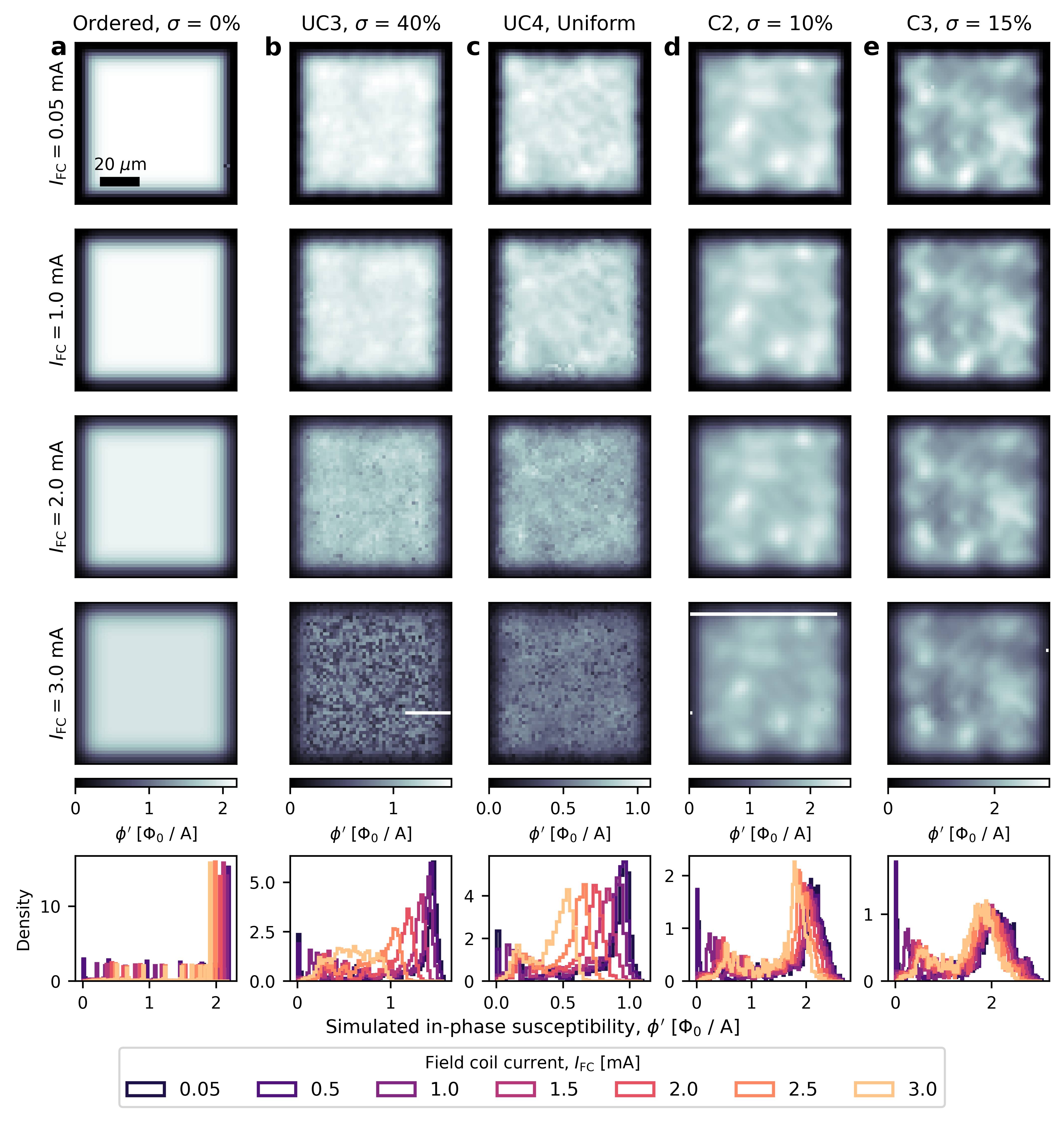}
	\caption{Inhomogeneous, nonlinear magnetic response in the junction network model. Each column shows the simulated in-phase susceptibility $\phi'$ for one of the arrays at three different applied field coil currents, $I_\mathrm{FC}$. The bottom row shows normalized histograms of $\phi'$ in the central $80\times80\,\mu\mathrm{m}$ region of each of the images, with a bin width of 0.025 $\Phi_0$/A. Note the difference in linearity between the uncorrelated arrays (columns b and c) and correlated arrays (columns d and e). The few rows with missing pixels in the images for $I_\mathrm{FC}=3.0\,\mathrm{mA}$ indicate simulations that timed out prior to completion. This figure represents a spatially-resolved version of Figure 5 (k-o) in the main text.}
	\label{fig:djn-nonlinear}
\end{figure*}

\begin{figure*}
	\centering
	\includegraphics[width=0.9\linewidth]{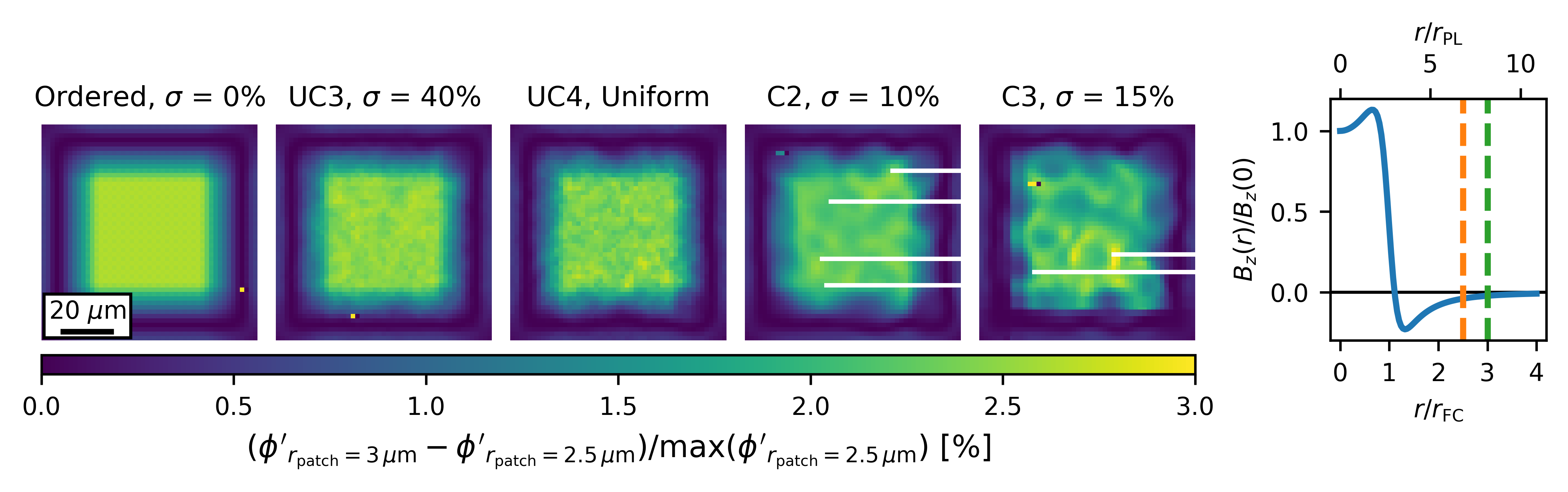}
	\caption{For all five arrays, increasing $r_\mathrm{patch}$ by 20\%, from $2.5\times r_\mathrm{FC}=17\,\mu\mathrm{m}$ (dashed orange line) to $3\times r_\mathrm{FC}=20.4\,\mu\mathrm{m}$ (dashed green line), increases the simulated susceptibility $\phi'$ by $<3\%$ of the maximum susceptibility and does not significantly change the spatial structure. For each array we use a junction cutoff radius $r_\mathrm{cutoff}=0.9\,\mu\mathrm{m}$. The rightmost panel shows the $z$-component of the magnetic field from the field located at $z_\mathrm{FC}=2\,\mu\mathrm{m}$ as a function of the radial distance $r$ from the center of the field coil, normalized by the field at $r=0\,\mu\mathrm{m}$. The point at which the field changes sign $r\approx 7.5\,\mu\mathrm{m}\approx1.1\times r_\mathrm{FC}\approx3\times r_\mathrm{PL}$ is the quantity $\rho_0$ discussed in the main text. The rows with missing pixels in the images for C2 and C3 represent simulations with $r_\mathrm{patch}=3\times r_\mathrm{FC}$ that timed out before completion due to the size of the NLP problem.
}
	\label{fig:djn-patch}
\end{figure*}

The discrete junction network (DJN) model exhibits nonlinearity, namely decreasing diamagnetic susceptibility with increasing applied field, as the gauge-invariant phase $\theta$ across junctions in the network becomes large enough that $\sin\theta\approx\theta$ is not a good approximation. Figure~\ref{fig:djn-nonlinear} demonstrates this behavior for all five of the arrays. At $I_\mathrm{FC}=2.5\,\mathrm{mA}$, the maximum flux through a $500\times 500\,\mathrm{nm}^2$ plaquette in the ordered array (column (a)) due to a field coil with radius $r_\mathrm{FC}=6.8\,\mu\mathrm{m}$ is $\approx0.028\,\Phi_0$, corresponding to a gauge-invariant phase of $\theta=2\pi\times0.028/4\approx0.044$ radians across each of the four junctions in the plaquette. Figure~\ref{fig:djn-nonlinear} shows that the nonlinearity is evident in the DJN model at applied fields far below one-quarter flux quantum per plaquette. Still, the onset of nonlinearity observed in experiment occurs at a much lower applied field than in the DJN model, as discussed in the main text.

\begin{figure}
	\centering
	\includegraphics[width=\linewidth]{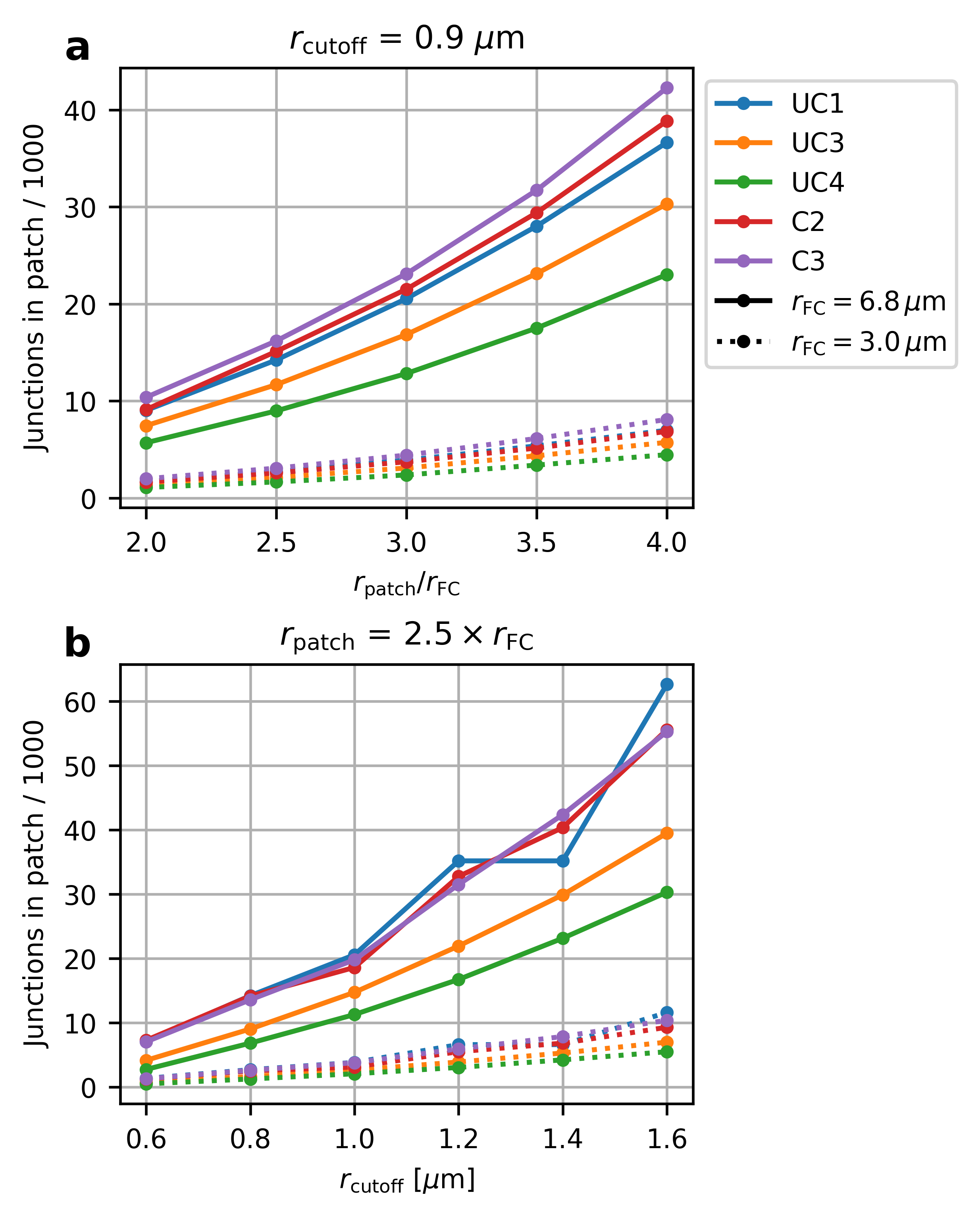}
	\caption{Scaling of the junction network model with field coil radius $r_\mathrm{FC}$, patch radius $r_\mathrm{patch}$, and junction cutoff radius $r_\mathrm{cutoff}$ for each of the five arrays studied here. (a) Total number of junctions (in thousands) in the network as a function of $r_\mathrm{patch}/r_\mathrm{FC}$ for $r_\mathrm{FC}=6.8\,\mu\mathrm{m}$ (solid lines) and $r_\mathrm{FC}=3.0\,\mu\mathrm{m}$ (dashed lines), assuming $r_\mathrm{cutoff}=0.9\,\mu\mathrm{m}$ . (b) Total number of junctions (in thousands) in the network as a function of $r_\mathrm{cutoff}$ for $r_\mathrm{FC}=6.8\,\mu\mathrm{m}$ (solid lines) and $r_\mathrm{FC}=3.0\,\mu\mathrm{m}$ (dashed lines), assuming $r_\mathrm{patch}=2.5\times r_\mathrm{FC}$ .}
	\label{fig:djn-scaling}
\end{figure}

\subsection{Scaling}
\label{sec:djn-scaling}
The junction network model scales poorly with the ratio of the field coil radius $r_\mathrm{FC}$ to the typical center-to-center island spacing. Given our field coil radius $r_\mathrm{FC}=6.8\,\mu\mathrm{m}$, for the ordered array UC1 with lattice constant $a=500\,\mathrm{nm}$, assuming a junction cutoff radius $r_\mathrm{cutoff}=0.9\,\mu\mathrm{m}$ and patch radius $r_\mathrm{patch}=17\,\mu\mathrm{m}=2.5\times r_\mathrm{FC}$, the total number of islands in the patch is $n\approx3,600$ and the total number of junctions is $m\approx14,200$. With such a large $r_\mathrm{FC}$, the number of junctions in the network (and therefore the number of variables and constraints in the NLP problem) grows very quickly with both $r_\mathrm{patch}$ and $r_\mathrm{cutoff}$ (see Figure~\ref{fig:djn-scaling}). Future measurements using SQUID susceptometers with a smaller field coil radius $r_\mathrm{FC}$~\cite{kirtley_scanning_2016} will be much less computationally costly to simulate, allowing us to explore these parameters of the DJN model in more detail.

\section{Continuous linear film model}

Here we describe an alternative model of the system that we developed, which did not reproduce the measured susceptibility as effectively as the junction network model (DJN). In this model, which we call the continuous linear film (CLF) model, we treat the system as a continuous 2D superconducting film with inhomogeneous effective magnetic penetration depth $\Lambda(x, y)=\lambda^2(x, y) / t$ whose magnetic response is governed by a generalized 2D London equation~\cite{Cave1986-js,Kogan2011-zn}:
\begin{equation}
\label{eq:london_2d}
\mathbf{H}(x, y)=-\mathbf{\nabla}\times[\Lambda(x, y)\mathbf{J}(x, y)],
\end{equation}
where $\lambda$ is the London penetration depth, $t\ll\lambda$ is the film thickness, $\mathbf{H}$ is the magnetic field in the film, $\mathbf{J}$ is the sheet current density in the film, and $\mathbf{\nabla}=\left(\frac{\partial}{\partial x}, \frac{\partial}{\partial y}\right)$.

\begin{figure*}
	\centering
	\includegraphics[width=\linewidth]{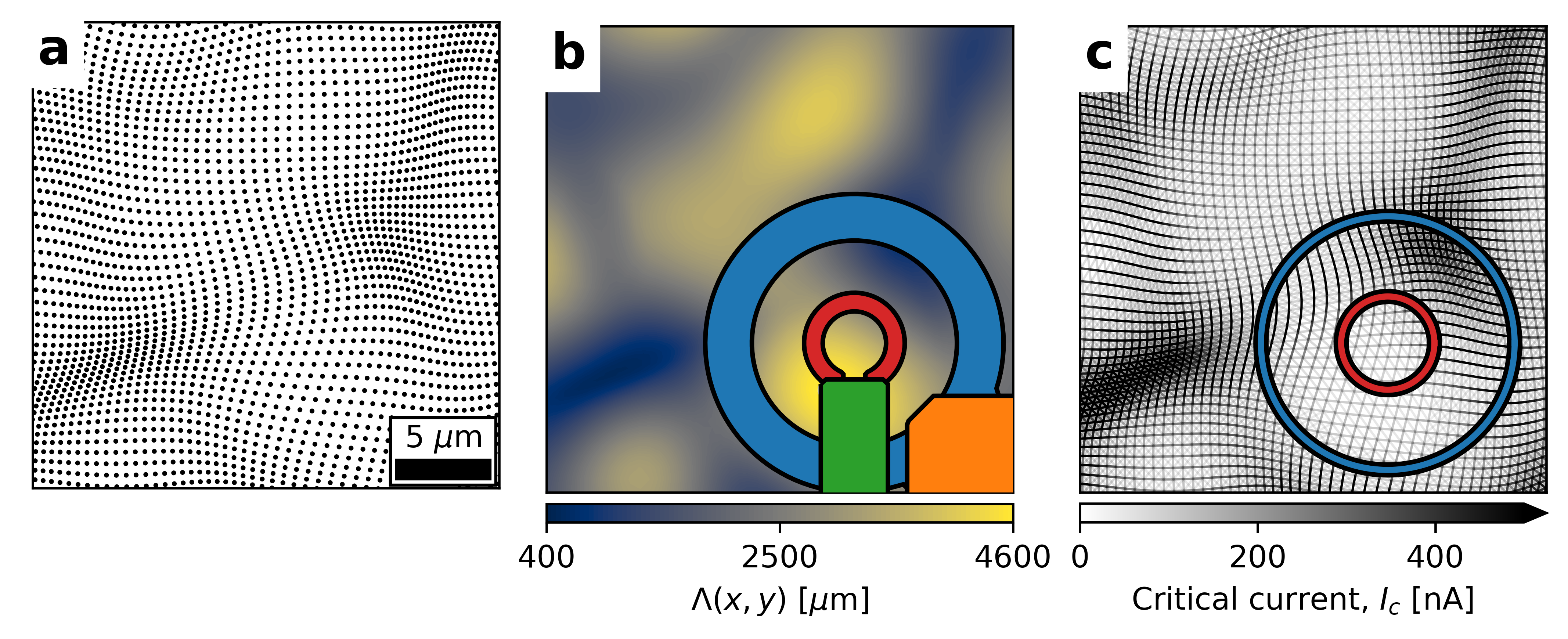}
  \caption{Schematic depiction of the CLF and DJN models. (a) Niobium island positions for a $25\times25\,\mu\mathrm{m}^2$ region of correlated array C3 ($\sigma=15\%$), with island diameters drawn to scale. (b) Effective penetration depth $\Lambda(x, y)$ for the same region, calculated using Equation~\ref{eq:Lambda_sim} assuming $I_0=260\,\mathrm{nA}$ with $d(x, y)$ obtained by averaging the local island spacing over 8 nearest neighbors. The colored polygons in the bottom right show the geometry of SQUID susceptometer simulated in the CLF model, again drawn to scale. (c) The network of Josephson junctions used to represent this region of the array in the DJN model. Each line represents a single junction, where we assume that niobium islands form junctions with all neighboring islands within a radius of $r_\mathrm{cutoff}=0.9\,\mu\mathrm{m}\approx2.1\times\xi_N$, where $\xi_N\approx0.425\,\mu\mathrm{m}$ is the normal metal coherence length. The shade of each line corresponds to the junction critical current, as shown by the colorbar. The blue circle shows the 1D loop with radius $r_\mathrm{FC}=6.8\,\mu\mathrm{m}$ used to model the field coil, and the red circle shows the 1D loop with radius $r_\mathrm{PL}=2.5\,\mu\mathrm{m}$ used to model the pickup loop. Regions in (a) that are more densely populated with islands have a shorter penetration depth in (b) and have a higher density of junctions in (c), and those junctions have larger critical currents.}
	\label{fig:clf-setup}
\end{figure*}

For an ordered Josephson junction array, the low-field effective penetration depth is given in S.I. units by $ \Lambda=\Phi_0/(2\pi\mu_0I_c)$, where $\mu_0$ is the vacuum permeability and $I_c$ is the critical current of each junction~\cite{tinkham2004introduction,phillips_influence_1993}. We therefore assume that the effective penetration depth of the film is given by
\begin{equation}
\label{eq:Lambda_sim}
\Lambda(x, y)=\frac{\Phi_0}{2\pi\mu_0I_c(x, y)}=\frac{\Phi_0}{2\pi\mu_0}\frac{1}{I_0}\left(\frac{d(x, y)}{d_0}\right)^2,
\end{equation}
where $d(x, y)$ is the ``local island spacing", $d_0=240\,\mathrm{nm}$ is the minimum edge-to-edge island spacing in the ordered array, and $I_0$ is a constant that corresponds to the maximum critical current per junction in the ordered array, which sets the overall strength of the screening in the system. To calculate $d(x, y)$, we take the as-designed island positions $(x_i, y_i)$ and, for each island $i$, calculate the average distance to the island's 8 nearest neighbors. This gives an estimate of $d(x_i, y_i)$ at each of the island positions, from which we compute $d(x, y)$ for any $(x, y)$ within the film via linear interpolation.

The arrays with correlated disorder (C2, C3) are locally ordered on a length scale given by the engineered correlation length, $\ell=5\,\mu\mathrm{m}$, so one might expect their local magnetic response at low applied fields to be approximately described by a local effective penetration depth $\Lambda(x, y)$, as given in Equation~\ref{eq:Lambda_sim}. In contrast, the arrays with uncorrelated disorder (UC3, UC4) are not ordered on any experimentally relevant length scale, so we do not expect Equation~\ref{eq:Lambda_sim} to capture their magnetic response.

\begin{figure*}
	\centering
	\includegraphics[width=0.9\linewidth]{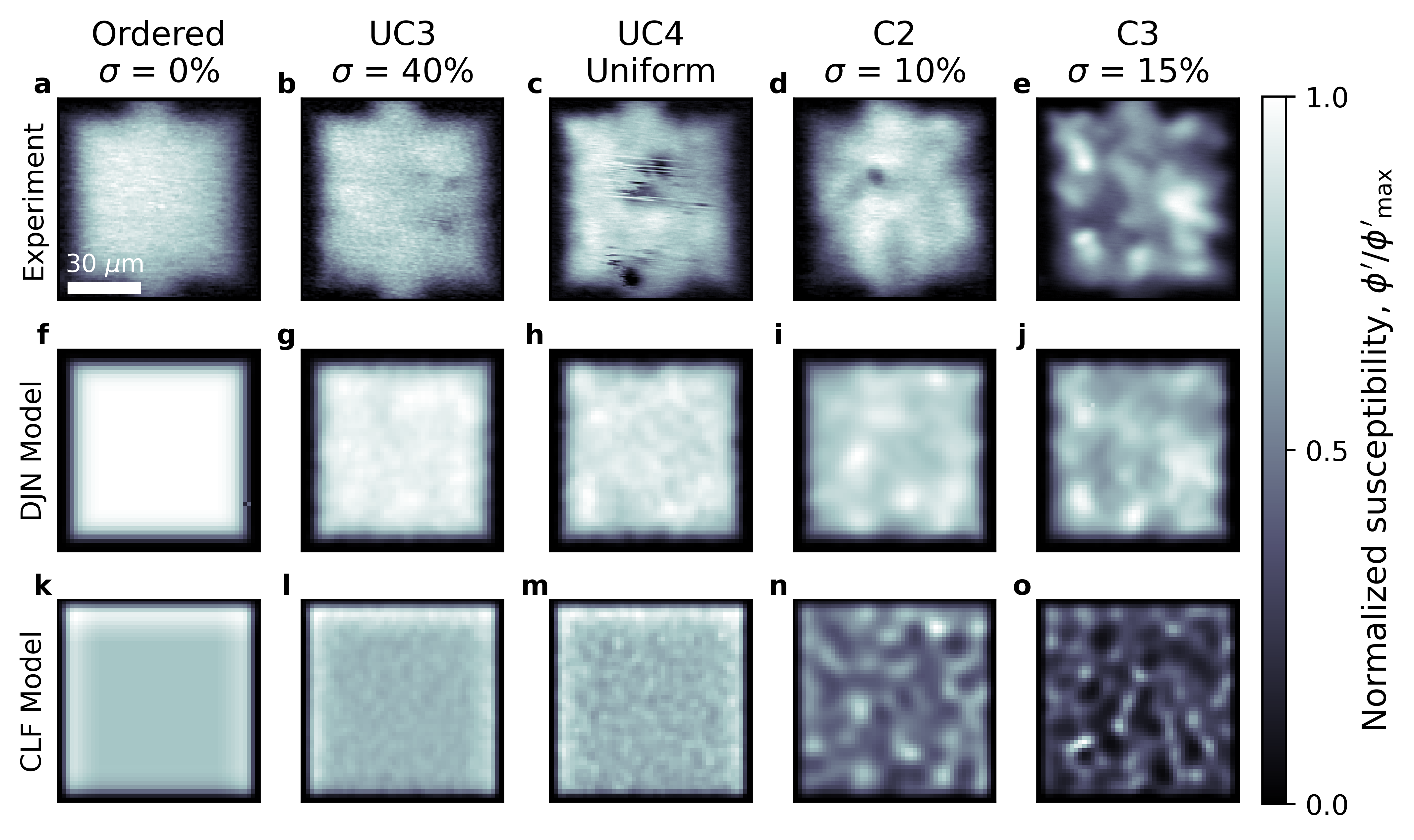}
	\caption{Comparison of results from the CLF and DJN models to the measured low-applied-field in-phase susceptibility $\phi'$ for all five arrays. Top row (a-e): Normalized in-phase susceptibility $\phi'/\phi'_\mathrm{max}$ measured at $I_\mathrm{FC}=0.05$ mA for panels (a, b, c, e) and $I_\mathrm{FC}=0.1$ mA for panel d. Middle row (f-j): $\phi'/\phi'_\mathrm{max}$ calculated using the DJN model with $r_\mathrm{cutoff}=0.9\,\mu\mathrm{m}$ and $r_\mathrm{patch}=17\,\mu\mathrm{m}=2.5\times r_\mathrm{FC}$. For both models we use a field coil current $I_\mathrm{FC}=0.05$ mA, a distance $z_\mathrm{FC}=2\,\mu\mathrm{m}$ between the sample surface and SQUID field coil, and a critical current scale of $I_0=260\,\mathrm{nA}$. Bottom row (k-o): $\phi'/\phi'_\mathrm{max}$ calculated using the CLF model, with local island spacing $d(x ,y)$ averaged over 8 nearest neighbors. Note that there are regions near the center of array UC4, with uniformly distributed island positions (panel c), where niobium islands have been scraped off due to contact with the SQUID susceptometer, which is not accounted for in either model.}
	\label{fig:clf-sim}
\end{figure*}

Having defined the film's effective penetration depth $\Lambda$, we model the film's response to the field due to a current $I_\mathrm{FC}$ flowing in the SQUID susceptometer field coil by self-consistently solving Equation~\ref{eq:london_2d} inside the film and the three superconducting layers of the SQUID, and Maxwell's equations in the vacuum regions between superconducting layers. We use the SuperScreen Python package~\cite{Bishop-Van_Horn2022-sy}, which implements a numerical method developed by Brandt~\cite{brandt_thin_2005} and first applied to scanning SQUID microscopy by Kirtley, \emph{et al.}~\cite{kirtley_scanning_2016,kirtley_response_2016} Just as in a scanning SQUID susceptometry measurement, the presence of the sample modifies the mutual inductance $\Phi_\mathrm{PL}/I_\mathrm{FC}$ between the SQUID field coil and pickup loop and this change in the mutual inductance, which depends upon $\Lambda(x, y)$ in the vicinity of the field coil, is the SQUID susceptibility signal. The simulated susceptibility is completely independent of $I_\mathrm{FC}$ due to the linearity of the London model (Equation~\ref{eq:london_2d}). See Figure~\ref{fig:clf-sim} for a comparison of results from the CLF and DJN models to the measured low-applied-field in-phase susceptibility $\phi'$ for all five arrays.

The DJN approach avoids two assumptions required by Equation~\ref{eq:Lambda_sim} and the CLF model, namely that the superconducting phase gradient across the array is small, and that the array is (at least locally) ordered~\cite{tinkham2004introduction}. Avoiding the first assumption allows us to explore nonlinearities in the magnetic response at large applied fields as the gauge-invariant phase $\theta$ across junctions in the network becomes large enough that $\sin\theta\approx\theta$ is not a good approximation. Avoiding the second assumption means that the model is applicable to arrays with both correlated and uncorrelated disorder.


\end{document}